\documentclass[twocolumn,showpacs,preprintnumbers,amsmath,amssymb]{revtex4}

\usepackage{amsmath,amssymb,amsfonts,latexsym}
\usepackage{graphicx}
\usepackage{dcolumn}
\usepackage{bm}

\begin{document}
\newcommand{\bra}[1]{\langle #1 |}
\newcommand{\ket}[1]{|#1\rangle}
\newcommand{\braket}[2]{\langle #1 | #2 \rangle}
\newcommand{\ketbra}[2]{|#1 \rangle \langle #2|}
\newcommand{\vel}[1]{\mathbf{\hat{#1}}}
\newcommand{\velm}[1]{\frac{d^3 \hat{#1}}{2 \hat{#1}^0}}

\newcommand{\abs}[2][]{\ensuremath{#1|#2#1|}}
\newcommand{\bk}[2]{\ensuremath{\langle #1|#2 \rangle}}
\newcommand{\bkb}[2]{\ensuremath{\langle #1|#2\}}}
\newcommand{\bks}[2]{\ensuremath{\langle #1|#2\succ}}
\newcommand{\kt}[1]{\ensuremath{|#1\rangle}}
\newcommand{\br}[1]{\ensuremath{\langle #1|}}
\newcommand{\kb}[2]{\ensuremath{| #1\rangle\langle #2|}}
\newcommand{\kp}[1]{\ensuremath{|#1)}}
\newcommand{\kts}[1]{\ensuremath{|#1\succ}}
\newcommand{\ktb}[1]{\ensuremath{|#1\}}} 

\newcommand{\cH}{\mathcal{H}}


\title{
Relativistic Resonances\\
 --- their Masses, Widths, Lifetimes, Superposition, and Causal Evolution
}

\author{Arno R. Bohm}
\email{bohm@physics.utexas.edu}
\author{Yoshihiro Sato}%
\email{satoyosh@physics.utexas.edu}
\affiliation{Department of Physics, University of Texas at Austin, Austin, Texas 78712-1081, USA
}%

\date{\today}

\begin{abstract}
	Whether one starts from the analytic S-matrix definition or the requirement of gauge-parameter independence in renormalization theory, a relativistic resonance is given by a pole at a complex value $s_R$ of the energy squared $s$.
	The complex number $s_R$ does not define the mass and the width separately, and the pole definition alone is also not sufficient to describe the interference of two or more Breit-Wigner resonances as observed in experiments.
	To achieve this and obtain a unified theory of relativistic resonances and decay, we invoke the decaying particle aspect of a resonance and associate to each pole a space of relativistic Gamow kets.
	The Gamow kets transform irreducibly under {\it causal} Poincar\'{e} transformations and have an exponential time evolution.
	Therefore one can choose of the many possible width parameters, the width $\Gamma_R$ of the relativistic resonance such that the lifetime $\tau=\hbar/\Gamma_R$.
	This leads to the parameterization $s_R=(M_R-i\Gamma_R/2)^2$ and uniquely defines these $(M_R, \Gamma_R)$ as the mass and width parameters for a resonance.
	Further it leads to the following new results: Two poles in the same partial wave are given by the sum of two Breit-Wigner amplitudes and by a superposition of two Gamow vectors with each Gamow vector corresponding to one Breit-Wigner amplitude.
	In addition to the sum of Breit-Wigner amplitudes the scattering amplitude contains a background amplitude representing direct production of the final state (contact terms).
	This contact amplitude is associated to a background vector representing the non-exponential energy continuum, omitting it gives the two interfering exponentials of the Weisskopf-Wigner methods.
	To accomplish all this required a minor modification in the foundation of quantum theory, which led to a quantum theory that contains the time asymmetry of causality.
\end{abstract}

\maketitle
\section{\label{sec:1}Introduction}


The Particle Data Group
\cite{ref:20}
lists two values for the mass and the width of $\Delta$ resonances.
	These two values differ from each other by 10 times the experimental error, one is called the Breit-Wigner mass and width and the other is called pole position.
	A similar situation holds for the $\rho$-meson.
For the mass and width of the $Z$-boson, the Particle Data Group
\cite{ref:20} 
  gives three definitions.
	When fitted to the line shape data of the same experiment 
\cite{ref:B66b, ref:3a}
the experimental values obtained for these three definitions differ from each other by about 10 times their experimental error.
	These examples indicate that one has problems with the understanding of resonances, in particular for relativistic resonances.
	This problem has its roots already in the foundations of quantum mechanics.

	The old quantum mechanics (based on the Hilbert space axiom including the use of Dirac kets) is a theory of stable states and reversible (unitary) time evolution.
	In contrast, quasistable states, like resonances in a scattering experiment or like decaying states in a decay experiment, are connected with an asymmetric or  ''irreversible'' time evolution 
\cite{ref:B3}.
	Thus they require a time asymmetric quantum theory, and in the absence of such a theory, their description can only be approximate and must contain some contradictions.
	If one is serious about Hilbert space mathematics one always runs into problems with the quantum theory of resonances and decaying states, basically because the vectors with exponential time evolution (as Gamow envisioned for quasistable states 
\cite{ref:B6}
) do not exist in the Hilbert space.
	In the heuristic treatment of scattering theory one just ignored the mathematical subtleties. 
	One worked with mathematically undefined kets  
\cite{ref:C1}, 
used $\pm i \epsilon$ to distinguish incoming from outgoing Lippmann-Schwinger kets 
\cite{ref:101}, 
and distinguished ``states at time $t'<t_0=$ time defined by preparation'' and ``states characteristic of the experiment'' observed at $t''>t_0$ 
\cite{ref:102}.
One restricted by fiat the time in $e^{iHt}$ to $t \geq 0$
\cite{ref:B52}, 
and for decaying states one postulated purely outgoing boundary conditions 
\cite{ref:103} 
undisturbed by the fact that this was in conflict with the unitary group evolution $-\infty<t<\infty$ which is a consequence of the Hilbert space axiom (Stone-von Neumann theorem 
\cite{ref:xx}).

	These heuristic methods were quite successful for physical applications, but when one compared it with mathematical consequences of the axioms in  
Ref.\ \onlinecite{ref:xx} 
one had contradictions.
	Examples of these are: the exponential catastrophe in which Gamow vectors and unitary time evolution conflicted; deviations from the exponential law
\cite{ref:104}; 
problems with (Einstein) causality 
\cite{ref:105}.

	In order to retain the empirically successful notions, like exponentially decaying Gamow states, the distinction between in- and out- Lippmann-Schwinger kets and between prepared states and detected observables, the Hilbert space axiom had to go.
	It was replaced by the Hardy space axiom which ascribes Hardy energy wave functions of conjugate analyticity to in-states and out-observables, respectively. 
	The use of Hardy energy wave functions then led to the desired association of a Breit-Wigner energy distribution to an exponentially evolving Gamow ket 
\cite{ref:28}, 
and also, unwittingly, to time asymmetry.

	In the {\it relativistic} case the Lippmann-Schwinger scattering states were always assumed to furnish a unitary (group) representations of the Poincar\'{e} transformations 
\cite{ref:33a}, 
just like the Dirac kets of the Wigner basis 
\cite{ref:30a},
despite their $\mp i \epsilon$ being in mathematical conflict with unitary group evolution.
	However, mathematically defined as Hardy space functionals, the Lippmann-Schwinger kets furnish only Poincar\'{e} semigroup representations into the forward (or backward) light cone, and this incorporates Einstein causality without requiring the separate axiom of local commutativity 
\cite{ref:cpt}.

	In quantum field theory there are no vectors corresponding to unstable states 
\cite{ref:106k}.
	Unstable states are eliminated from the set of asymptotic states by S-matrix unitarity 
\cite{ref:107}.
	They appear only as intermediators in some special forms of the propagator obtained by the Dyson summation formula 
\cite{ref:108}. 
	The precise form of the propagator depends upon the arbitrary choice of a renormalization point and so do the mass and width parameters defined by it.
	Though the complex pole definition of the $Z$-mass had been suggested as early as 1986 
\cite{ref:20a} 
the favored choice was the on-the-mass-shell definition. 
	This led to mass and width parameter which were gauge dependent in the next-to-the-next of the leading order 
\cite{ref:20b, ref:20c}.
	This gauge dependence disappeared when definitions based on the complex pole position of the propagators were employed 
\cite{ref:20a, ref:20b, ref:20c}.
	All this pointed to the definition of the resonance as the pole of the $j$-th partial S-matrix at a complex value $s_R$ of the center-of-mass scattering energy squared $s=(p_1+p_2)^2$.

	The position of the pole defines only the complex value $s_R$, not a mass $M$ and a width $\Gamma$ separately.
	How to split the complex number $s_R$ for the quasistable relativistic particle precisely into two real numbers of physical significance has not been completely agreed upon, except that the real part is predominantly connected with the mass and the imaginary part predominantly with the width or with the inverse lifetime.
	The inverse lifetime (which for the exponential decay law is equal to the initial decay rate) and the width are conceptually and experimentally different quantities; the former is measured using the exponential decay rate, the latter is measured as the width of a Breit-Wigner line shape.

	Whether one measures the width $\Gamma$ in an energy measurement or the lifetime $\tau$ in a time measurement is a question connected with the capabilities of the apparatuses, not a question related to the nature of the quasistable particle.
	For some relativistic particles it is possible to measure the width and for others the lifetime.
	One does not consider $\pi^0$ to be of different nature from $\eta$ because for $\pi^0$ one measures the lifetime and for $\eta$ one cannot.
	There exists no relativistic particle for which both width {\it and} lifetime have been measured.
	However for non-relativistic quasistable states one has an example for which both lifetime {\it and} width have been measured 
\cite{ref:25,ref:26},
so that the lifetime-width relation $\tau=\hbar/\Gamma$ could be tested and confirmed with high accuracy 
\cite{ref:x}.

	In the {\it non-relativistic} case one had a generally accepted heuristic method to relate width and lifetime, the Weisskopf-Wigner approximation 
\cite{ref:B20,ref:23}.
	For relativistic particles one should also like to define the lifetime and the width in such a way that the lifetime-width relation $\tau=\hbar /\Gamma$ holds.
	But the prevalent opinion in particle physics is that relativistic resonances are  complicated phenomena which cannot be defined by two real parameters, such as a mass $M$ and a width $\Gamma$.
	For instance, the $Z$-boson lineshape was 
considered as a Breit-Wigner amplitude with running width $\Gamma_Z(s)$ 
\cite{ref:20, ref:20a, ref:20c, ref:20b}.
	The same formula has also been used for hadrons 
\cite{ref:20}.

	After one noticed the problems with gauge invariance of the on-the-mass-shell definition 
\cite{ref:20b, ref:20c}, 
one became aware of the arbitrariness in the definition of the Z-boson mass and width, and concluded that there was no fundamental criterion to define the mass and width separately 
\cite{ref:41b}.
	Similar problems were also pointed out for the nucleon 
\cite{ref:21, ref:22d}
and meson 
\cite{ref:22r} 
resonances.
	This triggered the development of a unified theory for relativistic resonances and decaying particles 
\cite{ref:B57}.
	Without the concept of lifetime, the mass and the width of a relativistic quasistable particle cannot be uniquely defined.
	The S-matrix pole alone is not sufficient, one also needs the particle aspect of the relativistic system. 
	For stable relativistic particles the particle aspect is brought in by the relativistic quantum fields or equivalently 
\cite{ref:33a} 
by the representations of the Poincar\'{e} group 
of space-time transformations 
\cite{ref:30a}.

	Since the decay of a prepared state is believed to be a time asymmetric process 
\cite{ref:B3} 
and  the non-relativistic theory 
\cite{ref:28} 
required a semigroup, one expects that relativistic decaying states are also time-asymmetric and need, in place of Wigner's unitary Poincar\'{e} group representations, semigroup representations in the light cone.
	Semigroups are foreign to the traditional quantum theory in Hilbert space because with the Hilbert space boundary conditions the dynamical equations integrate always to a unitary group evolution 
\cite{ref:xx}.
	Nevertheless, long time ago, Schulman 
\cite{ref:35} 
gave a classification of Poincar\'{e} semigroup representations and even earmarked one of these classes $III\,\,E$, for the relativistic unstable particles.

	The same semigroup representations, called minimally complex representations because their momenta are given by $p_\mu=\sqrt{s_R}\hat{p}_\mu$, with real $\hat{p}_\mu$ (four-velocities) 
\cite{ref:33}, 
were obtained from the pole of the $j$-th partial S-matrix 
\cite{ref:34}.
	Therewith a resonance pole of $S_j(s)$ at $s_R$ was associated to a representation space of the causal transformations of relativistic spacetime.
	This representation space --- like the resonance pole characterized by $[s_R, j]$ --- is the space of a (single) relativistic resonance.
	The vectors in this space $[s_R,j]$ are the relativistic Gamow vectors $\psi^G_{[s_R,j]}$ which have exponential time evolution, as will be discussed in Sec.\,\ref{sec:4}, unburdened by the mathematical ballast of 
Ref.\ \onlinecite{ref:34}.
	Going beyond the results for a single relativistic resonance we then show in Sec.\,\ref{sec:4} how the relativistic Gamow kets provide all the properties that one observes for resonances and decaying states: interference of decaying states, superposition of resonance amplitudes; exponential decay for the resonance per se, and deviations from the exponential decay due to the non-resonant background amplitude.
	As a preparation for the relativistic theory in Sec.\,\ref{sec:4}, we give in Sec.\,\ref{sec:3} a brief review of the non-relativistic theory \cite{ref:28, ref:6} for which the Weisskopf-Wigner methods \cite{ref:B20, ref:23} serve as the starting point.
	The modifications needed in the foundations of quantum mechanics are most readily appreciated for the non-relativistic case of Sec.\,\ref{sec:3}.
	The relativistic concepts in Sec.\,\ref{sec:4} are introduced in analogy to Sec.\,\ref{sec:3} and on the basis of the phenomenological results of Sec.\,\ref{sec:2}.

	We consider in this paper mainly resonance formation
\begin{align} 
a+b \longrightarrow R \longrightarrow c+d.
\end{align}
	Resonance bumps, suggesting a Breit-Wigner amplitude, are also observed in resonance production 
\begin{align}
a+b \longrightarrow c+R, \quad R \longrightarrow e+f
\end{align}
	The relativistic Gamow vectors must therefore also emerge from the resonance production amplitude.
	This has indeed been shown \cite{ref:production1} and will be mentioned briefly below.
	The details are the subject of a separate publication \cite{ref:production2}.

\section{\label{sec:2}
Pole of the S-matrix versus Propagator Definition 
--- Two Different Values for Mass and Width}

	In the non-relativistic case the Lorentzian as a function of energy $E$ was the prominent choice (in nuclear and atomic physics) for the scattering amplitude of the resonating partial wave with angular momentum $j$:
\begin{eqnarray}
a_j^{BW}(E)=\frac{r^\eta}{E-z_R}, \quad z_R \equiv E_R-i\Gamma/2.
\label{eq:bwamp}
\end{eqnarray}
where $r^\eta$ is a constant.
	From this one conjectured the resonance amplitude for the relativistic hadron (e.g., $\pi N$) resonances by the following substitution
\begin{align}
&\mbox{Energy }E \longrightarrow W=\sqrt{s},
\nonumber 
\\
&\mbox{Resonance Parameters }(E_R,\Gamma) \longrightarrow (M,\Gamma). \nonumber
\end{align}
	Then one obtains for the resonance amplitude in the center-of-mass energy $W=\sqrt{s}$
\begin{eqnarray}
a_j^{BW}(W)=\frac{r^\eta}{W-W_P}, 
\quad
W_{P}\equiv M-i\Gamma/2.
\label{eq:bwamp-w}
\end{eqnarray}
	This defines the meaning of mass $M$ and width $\Gamma$ of a resonance.
	The resonance parameters $M$ and $\Gamma$ were determined by a fit of the resonance amplitude 
\eqref{eq:bwamp-w} 
with a slowly varying background $B_j(s)$:
\begin{eqnarray}
a_j(W) = a_j^{BW}(W) + B_j(W),
\label{eq:scattamp-w}
\end{eqnarray}
to the experimental data using cross sections $\sigma_j(W) \sim |a_j(W)|^2$, Argand diagrams
\cite{ref:5,ref:6,ref:14,ref:15}, 
and speed plots
\cite{ref:16}.

	The complex resonance parameter $W_P$ (and $z_R$) is associated with a pole of the S-matrix element $S_j(W)$ in the complex $W$-plane.
	Assuming analyticity of the S-matrix, except for a singularity due to the resonance at $W_P$, one can justify the amplitude 
\eqref{eq:scattamp-w} 
by the Laurent expansion if there is one pole.

	If there are two (or more) resonances in the same partial wave $j$, then a sum (superposition) of two (or more) resonance amplitudes was used:
\begin{eqnarray}
a_j(W) = \sum_{i=1}^2 a_j^{BW_i}(W)+B_j(W)
\label{eq:multires-scattamp-w}
\end{eqnarray}
where
\begin{eqnarray}
a_j^{BW_i}(W) = \frac{r^\eta_i}{W_{P_i}-W}, \quad W_{P_i}\equiv M_i-i\Gamma_i/2.
\label{eq:multires-bwamp-w}
\end{eqnarray}
This can no more be justified by the analyticity assumptions for the S-matrix using a Laurent expansions but it worked phenomenologically very well.

	Another starting point for the definition of a relativistic resonance is as the pole of the S-matrix $S_j(s)$ on the $s$-plane $(s=W^2)$.
	For the resonance amplitude $a_j^\mathcal{R}(s)$, one then takes
\begin{eqnarray}
a_j^\mathcal{R}(s) = a_j^{BW}(s) \equiv \frac{r}{s-s_R}=\frac{r}{s-\bar{M}_Z^2+i\bar{M}_Z \bar{\Gamma}_Z},
\label{eq:relbwamp}
\end{eqnarray}
which we call the relativistic Breit-Wigner amplitude with constant width.
	The complex number $s_R$ is the position of the pole on the second (or higher) Riemann sheet of the S-matrix, $r$ is a constant, the residue, and $(\bar{M}_Z, \bar{\Gamma}_Z)$ is one of many possible parameterizations of $s_R$ in terms of real numbers given by \eqref{eq:masswidth-a} below.
	This resonance amplitude \eqref{eq:relbwamp} is therefore also called the pole definition of a relativistic resonance 
\cite{ref:10}.

	The complex position $s_R$ does not fix the definition of the real parameters, mass $M$ and width $\Gamma$, because there are many different parameterizations of the complex constant $s_R$ in terms of two real parameters which one could interpret as mass and width. 
	Three definitions of some historical value are $(m_1,\Gamma_1)$, ($\bar{M}_{Z},\bar{\Gamma}_{Z}$), and ($M_{R},\Gamma_{R}$) given by the parameterizations,
\begin{subequations}
\begin{align}
s_R&=\frac{m_1^2- i m_1 \Gamma_1}{1+(\Gamma_1/m_1)^2},
\label{eq:masswidth-c}
\\
s_{R}&=\bar{M}_{Z}^{2}-i\bar{M}_{Z}\bar{\Gamma}_{Z},
\label{eq:masswidth-a} 
\\
s_{R}&=(M_{R}-i\frac{\Gamma_{R}}{2})^2. \qquad \qquad 
\label{eq:masswidth-b}
\end{align}
\label{eq:masswidth}
\end{subequations}
	There could be many other parameterizations. 
	We shall see in Sec.\,\ref{sec:4} as one of our main results that the relativistic transformation properties select one of these three parameterizations.

	The definition 
\eqref{eq:bwamp-w} 
by a pole on the $W$-plane and the definition 
\eqref{eq:relbwamp} 
by a pole on the $s=W^2$-plane are very similar.
	A pole of $S_j^{n'n}(s)$ in the second sheet of the $s$-plane at $s=s_R=(M_R-i\Gamma_R/2)^2$ is always connected with a pair of poles in the $W$-plane at $W_P=M_R-i\Gamma_R/2$ and at $W_P=-(M_R-i\Gamma_R/2)$, because
\begin{align*}
\frac{1}{s-s_{R}}
&=\frac{1}{W^2-(M_{R}-i\Gamma_{R}/2)^2} \notag \\
&=\frac{1}{W+(M_{R}-i\Gamma_{R}/2)}\frac{1}{W-(M_{R}-i\Gamma_{R}/2)}.
\end{align*}
	Since the pole at $W_P=-(M_R-i\Gamma_R/2)$ is far away from the physical region $(m_1+m_2)<W<\infty$, one obtains for $\Gamma_R/M_R \ll 1$: 
\begin{eqnarray}
\frac{1}{s-(M_R-i\Gamma_R/2)^2}
\approx \frac{1}{(W+M_R)}\frac{1}{(W-W_P)}, 
\label{eq:swrelation}
\end{eqnarray}
where
\begin{eqnarray*}
W_P=M_R-i\Gamma_R/2.
\end{eqnarray*}
	Therefore a fit using the relativistic Breit-Wigner amplitude 
\eqref{eq:relbwamp} 
with the parameters of 
Eq.\ \eqref{eq:masswidth-b} 
will lead to essentially the same values for the parameters $(M_R,\Gamma_R)$ as a fit of the resonance amplitude 
\eqref{eq:bwamp-w} 
for the parameters $(M,\Gamma)$: $M=M_R$ and $\Gamma=\Gamma_R$.
	We shall therefore restrict ourselves here to the pole definition 
\eqref{eq:relbwamp} 
and the relativistic S-matrix as analytic function of $s$.

	The most common parameterization of the $Z$-resonance amplitude is however not the relativistic Breit-Wigner amplitude \eqref{eq:relbwamp} but the resonance scattering amplitude with a mass $M_Z$ and an energy dependent width $\Gamma_Z(s)$, given by
\begin{align}
a_{j}^{\mathcal{R}}(s)
&= a_{j}^{om}(s)
= \frac{-\sqrt{s}\sqrt{\Gamma_{e}(s)\Gamma_{f}(s)}}{s-M_{Z}^{2}+i\sqrt{s}\Gamma_{Z}(s)}
\label{eq:omamp_a}\quad\\
\notag \\
&\approx \frac{-M_{Z}B_{e}B_{f}\Gamma_{Z}}{s-M_{Z}^{2}+i\frac{s}{M_{Z}}\Gamma_{Z}}
=\frac{R_{Z}}{s-M_{Z}^{2}+i\frac{s}{M_{Z}}\Gamma_{Z}},
\label{eq:omamp}
\end{align}
where $\Gamma_Z =\Gamma_Z(s=M_Z^2)$. 
	The function \eqref{eq:omamp} is the expression and notation used in most analyses of the experimental data for the $Z$-boson 
\cite{ref:20, ref:B66b, ref:3a}.
	It initially emerged from the on-mass-shell renormalization scheme with a ``natural choice'' for the on-shell mass and width 
\cite{ref:20b, ref:20c}.
	Theoretical arguments had led to the conclusion that in the $Z$-boson case the on-shell mass is gauge dependent in $\mathcal{O}(g^4)$ and the gauge dependence was shown to disappear when a definition based on the complex valued position of the propagator pole \cite{ref:20a} was employed 
\cite{ref:20b, ref:20c}.

	The renormalization theoretic definition of the parameters associated to resonances and decaying states is a delicate matter because it does not use only perturbation theory to a particular finite order, but it also involves Dyson summation of an infinite number of diagrams.
	Thus on the one hand it treats unstable particles like asymptotic states and on the other hand it uses infinite sums.
	On top of this it imposes an arbitrary renormalization condition.
	This lead Sirlin \cite{ref:20d} and Passera \cite{ref:20e} to the conclusion that the conventional on-shell definition is a problematic treatment of unstable particles.
	Also, from experience in non-relativistic quantum mechanics one knows that the decaying states (e.g., square well potential, or Auger states of $He$ \cite{ref:6}) are not obtained by starting from asymptotically free states, but by starting from bound states as the zeroth approximation.
	Decaying states and resonances are more similar to bound states than to interacting scattering states; the latter are connected asymptotically (Lippmann-Schwinger equation) to the interaction free continuum states.
	Therefore the pole definition, --- on the real axis for bound states and at complex energies for unstable states --- is much more natural, and it is comforting that the requirement of gauge invariance also leads to the complex pole for a relativistic resonance.

	Initially a correspondence between pole definition and gauge invariance was not compelling.
	One had assumed that the complex pole definition must be gauge invariant just because it is connected with the S-matrix pole.
	The gauge dependence of the on-shell mass was then proven by showing that the relation between the on-shell and the pole mass contained gauge dependent expressions.
	The formal proof that the complex pole mass is indeed gauge independent was only recently given, at the two-loop level \cite{ref:20g} and to all orders in perturbation theory \cite{ref:20f}.
	This then established another justification for making the complex pole at $s_R$ the starting point for the definition of a resonance and its state vector, as we shall do in Sec.\,\ref{sec:4}.

	The choice \eqref{eq:masswidth-c} for the parameterization for $s_R$ \cite{ref:20b} is the most practical one since with Eq.\ \eqref{eq:masswidth-c} one obtains
\begin{align}
\frac{r}{s-s_R}
&=
\frac{r}{s-\frac{m_1^2- i m_1 \Gamma_1}{1+(\Gamma_1/m_1)^2}}
=
\frac{r(1+i \Gamma_1/m_1)}{s-m_1^2+i \frac{s}{m_1}\Gamma_1}.
\label{eq:sirlin1}
\end{align}
	With $m_1=M_Z$ and $\Gamma_1/m_1=\Gamma_Z/M_Z$, this has the same denominator as the r.h.s.\ of Eq.\ \eqref{eq:omamp}. 
	Since Eq.\ \eqref{eq:omamp} is the formula employed in most analyses of the LEP measurements a fit to the data using Eq.\ \eqref{eq:omamp} will therefore directly provide the values $m_1$ and $\Gamma_1$.
	The parameterization \eqref{eq:masswidth-a} has been the most popular parameterization if one uses a constant width; the parameterization \eqref{eq:masswidth-b} has also been mentioned \cite{ref:20c} but never been made use of.
	If one comes from analytic S-matrix theory \cite{ref:10}, the definition \eqref{eq:masswidth-b} of mass and width  is the natural, but not the only possible choice. 
	In Sec.\,\ref{sec:4} we shall introduce a new hypothesis from which the parameterization \eqref{eq:masswidth-b} will be derived.

	An experimental discrimination between different functions for the resonance amplitude like, e.g.,
Eqs.\ \eqref{eq:bwamp-w} and \eqref{eq:relbwamp} or Eqs.\  \eqref{eq:omamp} and \eqref{eq:sirlin1}, is impossible because there is always a background term $B_{j}(s)$.
	Even for {\it one} resonance in the partial wave the  amplitude consists of the two parts:
\begin{eqnarray}
a_{j}(s)=a_{j}^{\mathcal{R}}(s) +B_{j}(s).
\label{eq:amp}
\end{eqnarray}
	A small term like the second term in Eq.\ \eqref{eq:sirlin1}, $\Gamma_1/m_1<10^{-4}$, can also not be noticed.
	The resonance amplitude $a_{j}^{\mathcal{R}}(s)$ describes the ``part'' of the scattering that goes through resonance, e.g., 
\begin{eqnarray}
\pi\,p \rightarrow \Delta \rightarrow \pi\,p \quad {\rm or} \quad e^{+}\,e^{-} \rightarrow Z^{0} \rightarrow f\,f.
\label{resonances}
\end{eqnarray} 
	The background amplitude $B_{j}(s)$ describes the non-resonant part (the contact term of the propagator
\cite{ref:17}, 
or the empirical background of 
Ref.\ \onlinecite{ref:14}) 
of the reaction
\begin{eqnarray}
\pi\,p \rightarrow \pi\,p \quad {\rm or} \quad e^{+}\,e^{-} \rightarrow f\,f.
\label{nonresonant}
\end{eqnarray} 

	Because of this ever-present unknown background, the experimental line shape data, no matter how accurate they may be, cannot discriminate between the different resonance amplitudes.
	One can always write
\begin{eqnarray}
a_j^{om}(s)=a_j^{BW}(s) + B'_j(s),
\label{eq:om-scatt-amp}
\end{eqnarray}
with a small or slowly varying function $B'_j(s)$ which can be shifted into the background amplitude $B_j(s)$.
	Therefore fits of the lineshape data using	
Eq.\ \eqref{eq:amp} with Eq.\ \eqref{eq:relbwamp} 
and using 
Eq.\ \eqref{eq:amp} with Eq.\ \eqref{eq:omamp} 
will turn out to be equally good as we shall discuss shortly.
	But the use of the different resonance amplitudes 
Eq.\ \eqref{eq:relbwamp} or \eqref{eq:omamp} 
will lead to fitted values for the resonance parameters $(M_Z,\Gamma_Z)$, $(\bar{M}_Z,\bar{\Gamma}_Z)$, and $(M_R, \Gamma_R)$ which significantly differ from each other.
	The question therefore is which of these $(M,\Gamma)$ is the right mass and width?

	The parameters $(M_{R},\Gamma_{R})$ and $(\bar{M}_{Z},\bar{\Gamma}_{Z})$ come from the same complex pole value $s_R$ and are therefore algebraically related; they are just two different parameterization of the same function \eqref{eq:relbwamp} for the amplitude $a^{\mathcal{R}}_j(s)$ that describes the resonance per se:
\begin{subequations}
\begin{align}
\bar{M}_Z = M_R \sqrt{1-\frac{1}{4}(\Gamma_R/M_R)^2}\,\,\\
\bar{\Gamma}_{Z} = \Gamma_{R}/\sqrt{1-\frac{1}{4}(\Gamma_{R}/M_{R})^2}.
\end{align}
\label{eq:param-relation}
\end{subequations}

	In contrast the parameters $(M_Z,\Gamma_Z)$ and $(\bar{M}_Z,\bar{\Gamma}_Z)$ are defined by two {\it different} functions, Eqs.\ \eqref{eq:omamp} and \eqref{eq:relbwamp} respectively.
	Their values are therefore obtained by fitting the same experimental data to two different lineshape functions, one containing for the $Z$-boson resonance amplitude Eq.\ \eqref{eq:omamp} (or Eq.\ \eqref{eq:sirlin1}), and the other containing Eq.\ \eqref{eq:relbwamp} for the $Z$-resonance per se.

%
	For the lineshape fits one therefore uses the following two cross section (and forward-backward asymmetry) formulas \cite{ref:B66b, ref:3a}: The lineshape formula
\begin{eqnarray}
\sigma_{tot}^0(s) \sim  \left[ \frac{G}{s} + \frac{s \cdot R+(s-M_Z^2)\cdot J}{|s-M_Z^2+is\Gamma_Z/M_Z|^2} \right]
\label{eq:sigma-om}
\end{eqnarray}
contains Eq.\ \eqref{eq:omamp} (or Eq.\ \eqref{eq:sirlin1}), and the lineshape formula
\begin{align}
\sigma^0_{tot}(s)\sim& \left[\frac{g_f}{s} + \frac{j_f\cdot (s-\bar{M}_Z^2) + r_f \cdot s}{(s-\bar{M}_Z)^2 + \bar{M}_Z^2 \bar{\Gamma}_Z^2}\right],
\label{eq:sigma-pole}
\\
&\mbox{with $f=$ had, e, $\mu$, $\tau$.},
\nonumber
\end{align}
contains Eq.\ \eqref{eq:relbwamp}.
	In here the parameters (for each fermion $f$) $G$ and $g_f$ describe the photon exchange ($G$ and sometimes also $g_f$ are usually assumed to be known, $G\sim \alpha_{e.m.}^2(M_Z)$), $R$ and $r_f$ measure the $Z$-peak height describing the $Z$-exchange, and $J$ and $j_f$ describe the photon-$Z$-boson interference.

	The lineshapes \eqref{eq:sigma-om} and \eqref{eq:sigma-pole} are derived if one assumes that the amplitude is a superposition of a photon ``Breit-Wigner resonance'' and a $Z$-boson Breit-Wigner resonance, i.e., given by the sum of two pole terms,
\begin{eqnarray}
a_j(s)&\sim &\frac{1}{s+i\epsilon} + \frac{R}{s-s_R}+B(s)
\notag
\\
&\approx & \frac{1}{s+i\epsilon} + \frac{R}{s-s_R},
\label{eq:z-gamma}
\end{eqnarray}
with $s_R=\bar{M}_Z^2 - i \bar{M}_Z \Gamma_Z = (M_R-i\Gamma_R/2)^2$, if the back ground $B(s)$ is neglected \cite{ref:3a}.
	The superposition \eqref{eq:z-gamma} emerges naturally in standard perturbation theoretical treatment, but in standard S-matrix theory superpositions of two pole terms like Eq.\ \eqref{eq:z-gamma} are not possible.
	The well known $i\epsilon$ in the amplitude \eqref{eq:z-gamma} is an ingredient of our new Hardy space axioms \eqref{eq:newaxiom} and \eqref{eq:hardy-whole} which specifies that the in- and out-energy wave functions $\braket{^+s}{\phi^+}$ and $\overline{\braket{^-s}{\psi^-}}=\braket{\psi^-}{s^-}$ must be analytic in the {\it lower} $s$-plane.
	The same Hardy function property will also result in the superposition of the two pole terms in Eq.\ \eqref{eq:z-gamma}.
	In Sec.\ \ref{sec:4} we shall introduce a new Hardy space hypothesis and justify the superpositions of two Breit-Wigner amplitudes also in analytic S-matrix theory.

		From the fits of the cross section (and asymmetry) data to Eq.\ \eqref{eq:sigma-om} one obtains the values $(M_Z, \Gamma_Z)$ and from the fit to Eq.\ \eqref{eq:sigma-pole} one obtains the values $(\bar{M}_Z, \bar{\Gamma}_Z)$ (and the other parameters $r$ and $j$).
			These mass and width values are given in Table \ref{tb:zboson}.
\begin{table}
\begin{center}
\begin{ruledtabular}
	\begin{tabular}{c c}
	$M_{Z}=91.1875 \pm 0.0021 GeV$ & $\bar{M}_{Z}=91.1526\pm0.0023GeV$ \\
	$\Gamma_{Z}=2.4939 \pm 0.0024 GeV$ & $\bar{\Gamma}_{Z}=2.4945\pm0.0024GeV$	
\end{tabular}
\end{ruledtabular}
\end{center}
\caption[Z-boson-mass]{$Z$-boson mass and width. $(M_Z, \Gamma_Z)$ are the values obtained from a lineshape fit using Eq.\ \eqref{eq:sigma-om} based on Eq.\ \eqref{eq:omamp}, and $(\bar{M}_Z, \bar{\Gamma}_Z)$ are the values obtained from Eq.\ \eqref{eq:sigma-pole} based on Eq.\ \eqref{eq:relbwamp}. The values are averages of the results obtained by ALEPH, DELPHI, L3, and OPAL \cite{ref:B66b}.}
\label{tb:zboson}
\end{table}
		The difference between the values from Eqs.\ \eqref{eq:sigma-om} and \eqref{eq:sigma-pole} (calculated from the Table \ref{tb:zboson}) is
\begin{subequations}
\begin{align}
M_Z &-\bar{M}_Z = 0.0349 \pm 0.0044 GeV,\\ 
\Gamma_Z &-\bar{\Gamma}_Z = -0.0006 \pm 0.0048 GeV.
\end{align}
\end{subequations}
	This difference is significant as compared with the experimental errors $\sigma_{M_Z}= 0.0021GeV$, and therefore one may ask the question which of these $(M,\Gamma)$ one should use.

	The values of $M_R$ and $\Gamma_R$ can be directly calculated from the exact relation \eqref{eq:param-relation}:
\begin{subequations}
\begin{align}
M_R &= 91.1611 \pm 0.0023 GeV,\\
\Gamma_R &= 2.4943 \pm 0.0024 GeV.
\end{align}
\label{eq:mass-width-r}
\end{subequations}
	Therewith we have already three different values of mass and width of a relativistic resonance which present day experiments can discriminate, and it is timely to ask for a theoretical criterion that distinguishes the right definition of $M$ and $\Gamma$.

	As far as the lineshape or resonance amplitude is concerned $(M_R,\Gamma_R)$ and $(\bar{M}_Z,\bar{\Gamma}_Z)$ are equivalent, whereas $(M_Z,\Gamma_Z)$ and $(\bar{M}_Z,\bar{\Gamma}_Z)$ belong to different lineshapes.
	But one can also relate $(M_Z,\Gamma_Z)$ and $(\bar{M}_Z,\bar{\Gamma}_Z)$ to each other by identifying the position of the maxima, $s_M^{BW}$ and $s_M^{om}$, of the two functions $|a_j^{BW}(s)|^2$ and $|a_j^{om}(s)|^2$:
%
\begin{subequations}
\begin{align}
s_M^{BW}&=\mbox{(maximum position of $|a_j^{BW}(s)|^2$)}
\nonumber
\\
&=\bar{M}_Z^2,
\\
s_M^{om}&=\mbox{(maximum position of $|a_j^{om}(s)|^2$)}
\nonumber
\\
&=M_Z^2 (1+(\Gamma_Z/M_Z)^2)^{-1}.
\end{align}
\end{subequations}
	Though there is not compelling reason for it, one can align their maxima, $s_M^{BW}=s_M^{om}$.
	This leads to
%
\begin{align}
\bar{M}_Z&= M_Z (1+(\Gamma_Z/M_Z)^2)^{-1/2}
\nonumber
\\
&=M_Z-0.0341 GeV,
\label{eq:mass-formula}
\end{align}
%
	Then one can also identify the values of $a_j^{BW}(s_M^{BW})$ and $a_j^{om}(s_M^{om})$.
	This brings in the residues $r$ of Eq.\ \eqref{eq:relbwamp} and branching in $R_Z$ of Eq.\ \eqref{eq:omamp} and leads to further complications \cite{ref:grassi}.
	But if one sets $r=R_Z(1+i\Gamma_Z/M_Z)^{-1}$ one obtains the standard relation \cite{ref:20} 
\begin{align}
\bar{\Gamma}_Z=\Gamma_Z (1+(\Gamma_Z/M_Z)^2)^{1/2}.
\end{align}
	With these identifications Eq.\ \eqref{eq:sirlin1} is written as
\begin{align}
a_j^{BW}(s) &= \frac{R_Z (1+i\Gamma_Z/M_Z)^{-1} (1+i\Gamma_1/m_1)}{s-m_1^2+i \frac{s}{m_1}\Gamma_1}
\nonumber
\\
&\approx \frac{R_Z}{s-m_1^2+i \frac{s}{m_1}\Gamma_1}
\label{eq:sirlin3}
\end{align}
which, with $M_Z=m_1$ and $\Gamma_Z=\Gamma_1$, is the formula \eqref{eq:omamp}, used in the lineshape formula \eqref{eq:sigma-om}.

	The identification \eqref{eq:mass-formula} is often presented like the definition of one set of parameters $M_Z$ and $\Gamma_Z$ in terms of another set of parameters $\bar{M}_Z$ and $\bar{\Gamma}_Z$, like for the identity \eqref{eq:param-relation}
%
\footnote{In Ref.\ \onlinecite{ref:B66b} it is actually the values of $M_Z \equiv \bar{M}_Z + 0.0341GeV$, not the values of $\bar{M}_Z$ which are listed for the ``S-matrix fits'' and in Ref.\ \onlinecite{ref:B66b} one calls these shifted values $M_Z \equiv \bar{M}_Z+34.1MeV$, but not $\bar{M}_Z$ of Table \ref{tb:zboson}, the S-matrix parameters.}.
	But since $(M_Z, \Gamma_Z)$ and $(\bar{M}_Z, \bar{\Gamma}_Z)$ are obtained in two different fits to two different functions, \eqref{eq:sigma-om} and \eqref{eq:sigma-pole} respectively, the equality \eqref{eq:mass-formula} is really only an approximation valid in a (large) neighborhood of their identified maxima $s_M^{BW}$ and $s_M^{om}$.

	For the practical question, which $M$ and $\Gamma$ is the ``right'' definition of mass and width of a relativistic resonance, the different meaning of the equalities \eqref{eq:param-relation} and \eqref{eq:mass-formula} is of no importance.
	There are two reasons for which a fit to the lineshape (cross-section and asymmetries) cannot settle this question:
\begin{enumerate}
\item The presence of the background amplitude makes it impossible to empirically distinguish between two different functions for the amplitude of resonance per se --- cf.\ Eq.\ \eqref{eq:om-scatt-amp}.
\item For one and the same amplitude function one can have in principle many different parameterizations --- like in Eq.\ \eqref{eq:masswidth} for the function \eqref{eq:relbwamp}.
\end{enumerate}
	Accepting the presently favored pole definition of a resonance one is lead to the relativistic Breit-Wigner resonance amplitude \eqref{eq:relbwamp}.
	But to distinguish between the different parameterizations \eqref{eq:masswidth-c}, \eqref{eq:masswidth-a}, \eqref{eq:masswidth-b} and more, one requires yet another aspect than the lineshape. 
	In Sec.\ \ref{sec:4} we choose for this the particle aspect:the resonance per se is identified with an exponentially decaying relativistic state of lifetime $\tau=\hbar/\Gamma$.
	In Sec.\ \ref{sec:4} it will be shown that of all the possible width parameters $\Gamma$ only $\hbar/\Gamma_R$ can be the lifetime $\tau$ and the inverse decay rate.

	Presently there may be only one example (in Refs.\ \onlinecite{ref:25} and \onlinecite{ref:26}) for which the lifetime-width relation $\tau=\hbar/\Gamma$ has been tested beyond the accuracy expected of the Weisskopf-Wigner approximation.
	But the validity of the exponential law (for the decay probabilities and rates) is needed for the definition of the total and partial initial decay rates $R(t=0)=1/\tau$ and $R_\eta$ ($\eta$ labeling the decay channels), for the branching ratios $B_\eta=R_\eta/R$ and for the partial widths $\Gamma_\eta \equiv B_\eta \Gamma$. 
	These definitions and relations are used so extensively that --- just in order to assure their validity --- one should take the exponential time evolution as the defining property of a resonance state vector.
	That such a state vector is precisely associated with the resonance pole --- as we shall see in Sec.\ \ref{sec:4} --- is an additional point in favor of the pole definition.

	For the well measured hadron resonances, $\Delta(1232)$
and $\rho$,  
the state of affairs are similar to the $Z$-boson situation.
	This is shown in Table \ref{tb:hadrons}. 
\begin{table}
\begin{center}
\begin{ruledtabular}
	\begin{tabular}{ccc}
	$\Delta^{++}$ & $M_{\Delta}=1231.88 \pm 0.29MeV$ & $\bar{M}_{\Delta}=1212.50\pm 0.24MeV$ \\
	$ $ & $\Gamma_{\Delta}=109.07\pm 0.48 MeV$ & $\bar{\Gamma}_{\Delta}=97.37 \pm 0.42MeV$  \\
\hline
	$\rho$ & $M_{\rho}=768.1 \pm 0.5 MeV$ & $\bar{M}_{\rho}=757.5 \pm 1.5MeV$ \\
	$ $ & $\Gamma_{\rho}=151.5 \pm 1.2MeV$ & $\bar{\Gamma}_{\rho}=142.5 \pm 3.5MeV$
	\end{tabular}
\end{ruledtabular}
\end{center}
\caption[hadron]{Hadron masses and widths \cite{ref:22d, ref:22r}. $(M_\Delta, \Gamma_\Delta)$ and $(M_\rho, \Gamma_\rho)$ are the values of the parameters in Eq.\ \eqref{eq:omamp} and $(\bar{M}_\Delta, \bar{\Gamma}_\Delta)$ and $(\bar{M}_\rho, \bar{\Gamma}_\rho)$ are the values of Eq.\ \eqref{eq:relbwamp}.
}
\label{tb:hadrons}
\end{table}
 	The different values for $M_\Delta$ and $\bar{M}_\Delta$ are extracted from the {\it same} experimental data set \cite{ref:20, ref:21,  ref:22d} but using {\it different} functions, Eqs.\ \eqref{eq:omamp} and \eqref{eq:relbwamp} respectively, 
using {\it different definitions} for the resonance mass $M$ and the width $\Gamma$. 
	The fits to both functions \eqref{eq:omamp} and \eqref{eq:relbwamp} were comparably good (except for the background dependence, see below), and none of these two functions could be ruled out on phenomenological ground. 
	But they produce significantly different values for mass and width.
%
%
	For the $\Delta$ resonance the difference between the two pole values $\bar{M}_\Delta$ defined by Eq.\ \eqref{eq:masswidth-a} and $M_{\Delta R}$ defined by Eq.\ \eqref{eq:masswidth-b} is within the experimental errors.
	Therefore we did not list $M_{\Delta R}$ here.

	The values $(M_\rho,\Gamma_\rho)$ defined by 
Eq.\ \eqref{eq:omamp} 
and $(\bar{M}_\rho,\bar{\Gamma}_\rho)$ defined by 
Eq.\ \eqref{eq:relbwamp} with Eq.\ \eqref{eq:masswidth-a} 
have also been extracted from the same set of data 
\cite{ref:22r} 
and differ also by about 10 times the quoted error.
	In addition to Ref.\ \onlinecite{ref:22r} a precise determination of the $\rho$-mass has also been performed in 
Ref.\ \onlinecite{ref:22a} 
(using a different data set) and their value was given as $762.4\pm1.8MeV=M_{\rho R}$.
	This differs (though not significantly) from the value $\bar{M}_\rho = 757.5\pm 1.5 MeV$ of Ref.\ \onlinecite{ref:22r}. 
	However, the value given in Ref.\ \onlinecite{ref:22a} uses the definition $(M_{\rho R}, \Gamma_{\rho R})$ of the parameterization Eq.\ \eqref{eq:masswidth-b}.
	Using the exact relation \eqref{eq:param-relation} between $(\bar{M}_\rho, \bar{\Gamma}_\rho)$ and $(M_{\rho R}, \Gamma_{\rho R})$ one calculates from $M_{\rho R}$ of Ref.\ \onlinecite{ref:22a} the value $\bar{M}^{calc}_\rho = 758.9 \pm 1.8 MeV$ which is in perfect agreement with the value of Ref.\ \onlinecite{ref:22r} in Table \ref{tb:hadrons}.
	Thus the values for the $\rho$-mass obtained in Ref.\ \onlinecite{ref:22r} and in Ref.\ \onlinecite{ref:22a} are in perfect agreement.

	In these precise fits (radiative) corrections and interference terms,  $\rho-\omega$ interference similar to the $Z-\gamma$ interference in Eq.\ \eqref{eq:z-gamma}, had to be taken into account to obtain a satisfactory fit.
	This is clear evidence for the superposition of two Breit-Wigner amplitudes, which will be shown in Sec.\ \ref{sec:4} to be a consequence of the Hardy space hypothesis also in S-matrix theory.

	As mentioned above the fits to Eqs.\ \eqref{eq:relbwamp} and \eqref{eq:omamp} are equally good.
	There is however a phenomenological aspect in favor of the S-matrix values $(\bar{M}_\Delta,\bar{\Gamma}_\Delta)$ and $(\bar{M}_\rho,\bar{\Gamma}_\rho)$.
	For the fits of the $\Delta-$ and $\rho-$ data, in addition to the resonance amplitudes Eq.\ \eqref{eq:relbwamp} or Eq.\ \eqref{eq:omamp} one always needs the background term $B(s)$ \cite{ref:22d, ref:22r}.
	If one uses the resonance amplitude \eqref{eq:relbwamp} one can use the same $B(s)$ for all channels.
	But if one uses the amplitude \eqref{eq:omamp} then one needs different background functions for different channels.


	The main argument in favor is the S-matrix pole definition Eq.\ \eqref{eq:relbwamp} is theoretical:
	Since the complex pole of the propagator has now been found to be the only gauge parameter independent definition of the $Z$ and $W$-boson masses 
\cite{ref:20f, ref:20g}, 
the pole of the S-matrix has become the clear theoretical choice.
	The pole definition in the $s$-plane also agrees (using the parameterization 
\eqref{eq:masswidth-b}) 
with the meritorious pole definition in the $W$-plane \eqref{eq:bwamp-w} and with the non-relativistic Breit-Wigner definition.
	Our conclusion of the lineshape discussions therefore is that the relativistic Breit-Wigner amplitude
\eqref{eq:relbwamp} 
represents the resonance per se and the on-the-mass-shell amplitude 
\eqref{eq:omamp} 
describes the resonance with some background 
\eqref{eq:om-scatt-amp}.
	If one favors the S-matrix pole definition of a resonance
\cite{ref:21} 
then the values $(\bar{M}_{\Delta}, \bar{\Gamma}_{\Delta})$, and $(\bar{M}_{\rho},\bar{\Gamma}_{\rho})$ are parameters that characterize the resonance per se, and $(M_{\Delta}, \Gamma_{\Delta})$ and $(M_{\rho},\Gamma_{\rho})$ are parameters describing the resonance together with some background.
	By the same argument as used for the hadron resonances, the parameters $(M_Z, \Gamma_Z)$ describe the $Z$-boson resonance with some background and the pole parameters $(\bar{M}_Z, \bar{\Gamma}_Z)$ --- or equivalently  by Eq.\ \eqref{eq:param-relation} ---  the pole parameters $(M_R, \Gamma_R)$ characterize the $Z$-boson per se.

	This still does not answer the question: Which of the parameterizations \eqref{eq:masswidth-c}, \eqref{eq:masswidth-a}, \eqref{eq:masswidth-b}, or others, should be used to define the mass and the width of a relativistic resonance?
	This question will be decided in Sec.\,\ref{sec:4}.

	In summary, the phenomenology of relativistic resonances based upon the analytic S-matrix for hadrons, and the quantum field theory for gauge bosons point toward the definition of a quasistable relativistic particle by a pole at the complex value $s_R$ in the $s$-plane (second sheet) of the S-matrix element $S_j^{n'n}(s)$ with angular momentum $j$.
	One observes a resonance by its (Breit-Wigner) lineshape, however, the precise meaning of mass and width of a relativistic resonance can not be fixed by the analysis of lineshape alone.

	For quasistable relativistic particles with values of $\Gamma/M \lesssim 10^{-7}$ one measures lifetimes by the exponential law.
	One even considers superpositions of two exponentially decaying states, e.g., for the neutral $K$ meson  
\cite{ref:B19}. 
	But their theoretical description is neither relativistic nor within the boundaries of conventional quantum mechanics, because one just takes eigenvectors of an arbitrary non-hermitian energy matrix to obtain the exponential time evolution states, whereas a state vector of a relativistic particle should be connected with the zero-th component $P_0$ of the total momentum operator, and time evolution should be a part of the Poincar\'{e} transformations.
	Thus the question arises: if resonances and decaying relativistic particles are qualitatively the same what does the the complex eigenvalue of the Poincar\'{e} generator $P_0$ have to do with the pole position $s_R$? 
	This will also be discussed in Sec.\,\ref{sec:4}.

\section{\label{sec:3}Modifying one axiom for a Weisskopf-Wigner Theory of non relativistic resonances}

	In order to relate the lifetime to the width of the lineshape, one requires a unified theory of resonance scattering and decay.
	Such a theory will then also determine which one of the width parameters, e.g., $\bar{\Gamma}_Z$ of  
Eq.\ \eqref{eq:masswidth-a} 
or $\Gamma_R$ of
Eq.\ \eqref{eq:masswidth-b} 
deserves to be called the width of a resonance defined by the S-matrix pole at $s_R$.
	We will construct this relativistic theory in analogy to the non-relativistic case, here we give a brief review of the non-relativistic theory of which Weisskopf-Wigner methods are approximations.

	For non-relativistic resonances one had the Weisskopf-Wigner methods by which one derived the decay probability $\mathcal{P}_\mathcal{R}(t)$ of a prepared resonance state with Breit-Wigner width $\Gamma$
\cite{ref:23} 
with the result:
\begin{eqnarray}
\mathcal{P}_\mathcal{R}(t) \sim e^{-\Gamma t/\hbar}+\Gamma\times(\textrm{additional terms}).
\end{eqnarray}
	From this one concluded that, at least in the ``approximation'' $\Gamma$$\times$(additional terms)$\to$$0$, the lifetime $\tau$ of a resonance $(E_R,\Gamma)$ is given by $\tau=\hbar/\Gamma$.
	If the resonance $\mathcal{R}$ has several ways to decay (decay channels), $\mathcal{R} \longrightarrow \eta_1,\eta_2,\eta_3 \ldots$, then the probabilities $\mathcal{P}_\eta(t)$ to find the decay product $\eta$ and the probability to find $\mathcal{R}$ undecayed fulfill
\begin{align}
&\mathcal{P}_\mathcal{R}(t) + \sum_\eta \mathcal{P}_\eta(t) 
= 1,
\nonumber \\
&\frac{d\mathcal{P}_\mathcal{R}}{dt}(t)
=-\sum_\eta \frac{d\mathcal{P}_\eta}{dt}(t)
=-\sum_\eta R_\eta(t).
\end{align}

	The lifetime $\tau$ is measured by fitting the counting rate, $\frac{1}{N} \frac{\Delta N_\eta(t)}{\Delta t}$, for any decay product $\eta$ to an exponential for the partial decay rate $R_\eta(t)$ (the intensity of the $\eta$ emission as a function of time):
\begin{eqnarray}
\frac{1}{N} \frac{\Delta N_\eta(t_i)}{\Delta t_i}
\approx \frac{d\mathcal{P}_\eta}{dt}(t) \equiv R_\eta(t)
=R_\eta e^{-Rt}, 
\label{eq:explaw}
\end{eqnarray}
where
\begin{eqnarray*}
\quad R=\sum_\eta R_\eta(0),
\end{eqnarray*}
and $\Delta N_\eta(t_i)$ is the number of decay products $\eta$ registered by the $\eta$-detector during the time interval $\Delta t_i$ around the time $t_i$ 
\footnote{
If the exponential law is fulfilled then the lifetime $\tau$ (defined as the average lifetime of the ensemble of the $N$ decaying particles) is $\tau=\frac{1}{R}$ independently of any quantum theory.
	If the exponential law is not fulfilled then the initial decay rate is not necessarily the inverse lifetime.
}.

	This exponential law has been compared with observations for more than a century 
\cite{ref:xxxa}.
It has been confirmed for values of the decay rate $R$ over many orders of magnitude $(10^{-17}$ -- $10^{16})s^{-1}$ and some reported non-exponential behavior, e.g., Ref.\ \onlinecite{ref:wilkinson}, may be attributed to the background amplitude, Eqs.\ \eqref{eq:bg-pole-final-ez} and \eqref{eq:bg-pole-distribution2} below.
	The exponential law can also be justified by intuitively correct heuristic  arguments
\footnote{
If the number of decay products $\Delta N(t)=\Delta(\sum_\eta N_\eta(t))$ counted in the time interval $\Delta t$ is proportional to the number $N_\mathcal{R}(t)$ of decaying objects: $\Delta N(t)= R N_\mathcal{R}(t)\Delta t$ where $R$ is a {\it constant} in time then $N_\mathcal{R}(t)=N_\mathcal{R}(0) e^{-Rt}$.
}.
	Therefore the exponential law of 
Eq.\ \eqref{eq:explaw} 
for a spontaneously decaying state without background can be considered as one of the well established laws of physics.
	If a theory does not fulfill the exponential law one should not fault the exponential law 
\cite{ref:104} 
but the theory.

	The probabilities $\mathcal{P}_\eta(t)$ are in quantum theory given by Born probabilities.
	If the observable has the properties of the decay products $\eta$, described by a projection operator $\Lambda_\eta$, and the decaying state vector is described by $\phi^D(t)$, then the probability for $\Lambda_\eta$ in $\phi^D(t)$ is given by the Born probability:
\begin{eqnarray}
\mathcal{P}_\eta(t)
&=&Tr(\Lambda_\eta \ketbra{\phi^D(t)}{\phi^D(t)})
=|\braket{\psi_\eta}{\phi^D(t)}|^2 \nonumber\\
&=&|\braket{\psi_\eta(t)}{\phi^D}|^2,\quad 
\label{eq:bornprob}
\end{eqnarray}
for $\Lambda_\eta =\ketbra{\psi_\eta}{\psi_\eta}$.
	One can show that in the Hilbert space $\mathcal{H}$ of conventional quantum mechanics there exist no such state vector $\phi^D(t)$ for which the probabilities
\eqref{eq:bornprob} 
obey the exponential law
\footnote{
Under the standard assumption that $\Lambda_\eta$ are projection of positive operators in $\mathcal{H}$ or $\psi_\eta \in \mathcal{H}$, the Hamiltonian is self-adjoint and bounded from below.
}.
 	At best a Hilbert space vector can describe an exponentially decaying state {\it with} some background (like the scattering amplitudes of 
Eq.\ \eqref{eq:amp}, 
which in addition to the resonance amplitude has some background).
	 The problem is therefore again a problem of separating the quasistable state vector with exponential time evolution from the background; in the same way as the Breit-Wigner amplitude for the resonance per se 
\eqref{eq:relbwamp} 
had been separated from the rest of the scattering amplitude.

	We want a resonance and an exponentially decaying state to be just different appearances of one and the same physical object, the quasistable quantum state. 
	Then we have to associate the Breit-Wigner amplitude to a ket $\psi^G$:
\begin{eqnarray}
a_{j}^{BW}(E) = \frac{r^{\eta}}{E-(E_{R}-i\Gamma/2)} 
\Longleftrightarrow 
\psi^G
\label{eq:bw-exp-relation}
\end{eqnarray}
with the properties
\begin{align}
e^{-iHt}\psi^G &= e^{-iz_R t}\psi^G,
\label{eq:bw-exp-relation2}
\\
H \psi^{G}&= z_R \psi^G
\end{align}
with $z_{R}= E_{R}-i\Gamma/2$, and $\phi^D \in \mathcal{H}$ must be separated like 
Eq.\ \eqref{eq:amp} 
into 
\begin{eqnarray}
\phi^D = \psi^G+\phi^{bg}.
\label{eq:d-gv-bg}
\end{eqnarray}
	Such a vector $\psi^G$, which we will call Gamow vector
\cite{ref:B6}, 
cannot be a regular vector in the Hilbert space since the Hamiltonian is self-adjoint and semibounded.
	However it can be one of the generalized eigenvectors (kets), which are defined as (continuous anti-linear) functionals.

	Functionals $F(\psi)$ on a linear space $\Phi$ are mathematically defined by the properties:
\renewcommand{\labelenumi}{\theenumi.)}
\begin{enumerate}

\item
$F(\alpha \psi + \beta \phi) = \bar{\alpha} F(\psi) + \bar{\beta} F(\phi)$ for every $\phi, \psi \in \Phi$ ; $\alpha, \beta \in \mathbb{C}$ (antilinearity).

\item
$F(\phi_\nu) \rightarrow F(\phi)$ as $\nu \rightarrow \infty$ for every sequence
$\phi_\nu$ that converges in the space $\Phi$ to $\phi$: $\phi_\nu \rightarrow \phi$ as $\nu \rightarrow  \infty$ (continuity).

\end{enumerate}
	The set of functionals on a space $\Phi$ forms again a space denoted by $\Phi^\times$ and called the dual of $\Phi$.
	All functionals $f(\phi)$ on the Hilbert space $\phi \in \mathcal{H}$ are given by the scalar product with a vector of $\mathcal{H}$ which we call also $f$: $f(\phi)=(\phi,f)$; this means $\mathcal{H}^\times = \mathcal{H}$. 
	But if $\Phi$ is a ``nicer'' space (not represented by Lebesgue square integrable functions but by smooth rapidly decreasing functions, Schwartz space) than the set of functionals on $\Phi$, $\Phi^\times$ is larger than $\Phi$ and than $\mathcal{H}$.
	Thus one has a triplet of spaces
\begin{eqnarray}
\Phi \subset \mathcal{H} = \mathcal{H}^\times \subset \Phi^\times
\label{eq:rhs}
\end{eqnarray}
	As noted, in the 4-th edition, by Dirac 
\cite{ref:C1},
``ket vectors form a more general space than a Hilbert space''.
	They are elements of an extended space $\Phi^\times$.
	This extended space $\Phi^\times$ is determined (defined) by the choice of $\Phi$.
	The ``nicer'' the elements in $\Phi$, i.e., the smaller the subspace $\Phi$ of $\mathcal{H}$, the larger is the space $\Phi^\times$, and that means the ``weirder'' are the kets in the extended space $\Phi^\times$.
	The Dirac kets are eigenkets with real continuous eigenvalues of a self adjoint Hamiltonian $H$:
\begin{eqnarray}
H \ket{Ejj_3\eta} = E \ket{Ejj_3\eta}, \quad 0\leq E < \infty.
\label{eq:diracket}
\end{eqnarray}
	They can be mathematically defined (and have been defined 
\cite{ref:cpt}) 
as functionals on the Schwartz space $\Phi$.
	The precise meaning of 
Eq.\ \eqref{eq:diracket} 
for $\ket{Ejj_3\eta}\in \Phi^\times$ is then
\begin{align}
\braket{H \psi}{Ejj_3 \eta} \equiv \bra{\psi} H^\times \ket{Ejj_3\eta}=E \braket{\psi}{Ejj_3\eta} 
\nonumber
\\
\mbox{for all $\psi \in \Phi$},
\label{eq:diracket-space}
\end{align}
and $H^\times$ is the (unique) extension of the adjoint $H^\dagger=H$ to $\Phi^\times$.
	Every physical vector representing a state $\phi$ or an observable $\ketbra{\psi}{\psi}$ can be written according to the Dirac basis vector expansion (nuclear spectral theorem of $\Phi \subset \mathcal{H} \subset \Phi^\times$) in terms of the kets $\ket{Ejj_3\eta}$ as
\begin{eqnarray}
\psi = \sum_{j\eta} \int_0^\infty dE \ket{Ej\eta}\braket{Ej\eta}{\psi}
\equiv \int_0^\infty dE \ket{E}\braket{E}{\psi},
\label{eq:diracket-expansion}
\end{eqnarray}
where the notation on the very right suppresses the discrete quantum numbers $j,j_3, \eta$. 
	The components along the basis vectors $\ket{E}$ are the energy wave functions $\braket{Ej\eta}{\psi} \equiv \braket{E}{\psi}\equiv \psi(E)$ which for the abstract Schwartz space $\Phi$ are the Schwartz space functions $\mathcal{S}$ (smooth rapidly decreasing):
\begin{eqnarray}
\psi \in \Phi \Longleftrightarrow \braket{E}{\psi} \in \mathcal{S}|_{\mathbb{R}_+}.
\label{eq:schwartz-func}
\end{eqnarray}

	Conventionally it is assumed that the states $\phi$ as well as the observables $\psi$ in the Born probabilities like 
\eqref{eq:bornprob} 
are both in the same space $\Phi$: $\phi,\psi \in \Phi$.

	In quantum physics, however, one always distinguishes between states $\phi$ and observables $\ketbra{\psi}{\psi}$; state is what is prepared by a preparation apparatus (accelerator) but observable is what is detected by a registration apparatus (detector).
	The quantities that are measured (as counting ratios of detectors like in 
Eq.\ \eqref{eq:explaw}
) are the Born probabilities (or probability rates) to detect an observable $\psi$ in the state $\phi$:
\begin{eqnarray*}
\mbox{``Born probability for $\psi$ in $\phi$''} = |(\psi,\phi)|^2.
\end{eqnarray*}
	The hypothesis of conventional quantum mechanics
\footnote{
The Hilbert space axiom 
\cite{ref:xx} 
is even stronger: $\{\phi\}=\{\psi\}=\mathcal{H}$ and since $\mathcal{H}^\times=\mathcal{H}$ there are no kets in $\mathcal{H}^\times$ that are not already in $\mathcal{H}$.
} 
states that
\begin{eqnarray}
&&\mbox{the set of prepared state $\{\phi\}$}
\nonumber\\
&&=\mbox{the set of observables $\{\psi\}$} 
\nonumber 
\\
&&= \Phi.
\label{eq:stdqm}
\end{eqnarray}
	This does not account for the fact that experimentally the observables $\psi$ and the prepared states $\phi$ represent different physical entities, e.g., the $\phi$'s are associated to the accelerator and the $\psi$'s to the detector.

	Under the axiom 
\eqref{eq:stdqm} 
there is only one kind of kets, the $F \in \Phi^\times$.
	In contrast, scattering theory uses two kinds of kets representing in-coming and out-going plane and/or spherical waves.
	The eigenkets of the exact Hamiltonian $H=H_0+V$ in scattering theory are not ordinary Dirac kets, i.e., elements of the dual of the Schwartz space
        $\Phi^\times$, but they are kets which also have meaning for
        complex values $E \pm i \epsilon$ with infinitesimal $\epsilon
        >0$. In scattering theory, one uses two solutions of
        the eigenvalue equation for the same eigenvalue $E$:
        \begin{equation}
                \label{eq:eigen-ls}
                H \ket{Ejj_3\eta ^{\mp}} = E \ket{Ejj_3\eta ^{\mp}}, \,\, \ 0
        \leq E < \infty.
        \end{equation}
        Here the superscript $\mp$ refers to the $\mp i \epsilon$ in
        the denominator of the Lippmann-Schwinger (integral) equation:
\begin{eqnarray}
\ket{Ejj_3\eta^\mp}=\ket{Ejj_3\eta}+\frac{1}{E-H_0\mp i \epsilon}V\ket{Ejj_3\eta^\mp}.
\label{eq:ls}
\end{eqnarray}
	This indicates that $\ket{Ejj_3\eta ^\mp}$ must be continued from the real (physical) energies into the complex lower half plane for ($-$) and into the upper half  plane for ($+$). 
	This means the complex conjugate of the wave
        functions, $\overline{\braket{^\mp E}{\psi ^\mp}} = {\braket{\psi
         ^\mp}{E ^\mp}}$, must not only be smooth functions of $E$
        like in
Eq.\ \eqref{eq:schwartz-func}
	but they must also be functions
        that have an analytic continuation into the complex energy
        plane, in particular $\braket{\psi^-}{E ^-}$ and $\braket{^+E}{\phi^+}$ must have an analytic continuation into the lower half plane. 
	Hardy functions, elements of $\mathbf{H}^2_{\mp} \cap \mathcal{S}|_{\mathbb{R}_+}$, have this property; they are the boundary values from below ($-$) or above ($+$) of analytic functions in the half-planes $\mathbb{C}_\mp$
\footnote{
All that one needs to know about Hardy functions to follow this paper is, that Hardy functions are boundary values of analytic functions in the lower (or upper) complex semiplane which vanish sufficiently fast at infinity, and that the mathematical properties used in the derivations of the following sections are correct.
	For the definitions and other properties of Hardy class functions see Appendix of 
Ref.\ \onlinecite{ref:28} 
and references thereof.
}.
	We turn this into a precise mathematical hypothesis.

	The energy wave functions of a scattering system fulfill
\begin{equation}        
\label{hardy-wave} 
{\braket{\psi^\mp}{E ^\mp}} \in \mathbf{H}^2_{\mp} \cap \mathcal{S}|_{\mathbb{R}_+},
\end{equation}
and this implies $\braket{^\mp E}{\psi ^\mp} \in \mathbf{H}^2_{\pm} \cap \mathcal{S}|_{\mathbb{R}_+}$.
        The generalized eigenvectors
\eqref{eq:eigen-ls}
representing out ($-$) and in ($+$) solutions of the Lippmann-Schwinger equation 
\eqref{eq:ls} 
are therefore kets in two different Hardy Rigged Hilbert Spaces:
        \begin{equation}
        \Phi_\pm \subset \mathcal{H} \subset \Phi_\pm^\times, \ \ \ \
        \ \
        \ket{E j \eta ^\mp} \in \Phi_\pm^\times.
	  \label{eq:hrgs}
        \end{equation}

	Defined as functionals on the Hardy space, the Lippmann-Schwinger kets can be analytically continued into the entire complex half plane as long as there are no singularities in the way.
	The energy half planes that we shall choose are those of the second (or higher) sheet of the S-matrix element with angular momentum $j$, $S_j(E)$, since the resonance poles are on these sheets.

	The Gamow vector which we need for 
Eq.\ \eqref{eq:bw-exp-relation2} 
is not one of the analytically continued Lippmann-Schwinger kets $\ket{z^-}$ in the lower complex plane, because $z_R$ is a singular points (first order pole) of $S_j(z)$.
	It also fulfills slightly different (purely out-going boundary) condition from that of the Lippmann-Schwinger kets.
	But like the Lippmann-Schwinger kets the Gamow vector is also mathematically defined as a functional $\psi^G(\psi^-)\equiv\braket{\psi^-}{\psi^G}$ on the Hardy space $\{\psi^-\}=\Phi_+$ of out-observables $\psi^-$.
	These $\{\psi^-\}$ include the decay products $\{\psi^-_\eta\}$ of the decaying state but also the out particles of a (resonance) scattering experiment.
	In terms of the Lippmann-Schwinger kets the Gamow ket can be defined by 
\begin{align}
\frac{\braket{\psi^-}{\psi^G}}{\sqrt{2\pi \Gamma}}
&=
\braket{\psi^{-}}{z_{R}j j_3 \eta^-} \nonumber \\
&=
\frac{i}{2\pi}\int_{-\infty_{II}}^{+\infty}dE\,\, \frac{\braket{\psi^{-}}{Ejj_3\eta^{-}}}{E-z_{R}}
\label{eq:integral1} 
\end{align}
for all $\psi^{-}\in\Phi_{+}$, where $z_R=E_R-i\Gamma/2$ is the pole position of $S_j(z)$.
	Omitting the arbitrary $\psi^- \in \Phi_+$, 
Eq.\ \eqref{eq:integral1} 
can be written as an equation between functionals:
\begin{eqnarray}
\psi^G
\equiv
\int_{-\infty_{II}}^{+\infty}
dE \,\,\ket{Ejj_3\eta^-}\frac{\sqrt{\frac{\Gamma}{2\pi}}}{E-z_R}.
\label{eq:nonrel-gv}
\end{eqnarray}
	This expresses the Gamow ket as a continuous superposition of the Lippmann-Schwinger kets $\ket{Ejj_3\eta^-}$ similarly to the Dirac basis vector expansion 
\eqref{eq:diracket-expansion}.
	The energy wave function $\braket{Ejj_3\eta}{\psi^G}$ of $\psi^G$ is the non-relativistic Lorentzian \eqref{eq:bwamp} which, however, in Eq.\ \eqref{eq:nonrel-gv} 
extends along the whole real axis, with $-\infty_{II} <E \leq 0$ in the second sheet right below the real axis (denoted by $II$) and along the cut of the ``physical'' scattering energies $0\leq E<\infty$.

	The Gamow vector 
\eqref{eq:nonrel-gv} 
can be shown 
\cite{ref:28} 
to have the property that it is an eigenket of the self-adjoint Hamiltonian $\overline{H}$ (in the sense of 
Eq.\ \eqref{eq:diracket-space})  
with eigenvalue $z_R$
\footnote{
$\overline{H}$ is the self-adjoint closure of $H$, and $H^\times$ in 
Eqs.\ \eqref{eq:eigenvalue} and \eqref{eq:timeevolution} 
is the dual of $H$ which is the uniquely defined extension of $\overline{H}=H^\dagger$ to $\Phi_-^\times$ 
\cite{ref:29}.
}.
\begin{eqnarray}
\braket{H\psi_\eta^-}{\psi^G} 
&\equiv& 
\bra{\psi_\eta^-} H^\times \ket{\psi^G}
=
(E_{R}-i\Gamma/2) \braket{\psi_\eta^-}{\psi^G}
\label{eq:eigenvalue}
\end{eqnarray}
for all $\psi_\eta^- \in \Phi_+$.
	The vector $\psi^G \equiv \ket{E_R-i\Gamma_R/2, jj_3n^-}\sqrt{2\pi \Gamma} \,$ is a generalized eigenvector like the Dirac ket, but since it is a functional on the space of analytic (Hardy) functions it can also have complex eigenvalues of (essentially) self adjoint operators $H$.
%

	The in-state vectors $\phi^+$ and the out-observable vectors $\psi^-$ are given by the  expansions:
\begin{subequations}
\begin{align}
\Phi_-\ni\phi^+ &= \sum_{j_3\eta} \int_0^\infty dE \ket{Ejj_3\eta^+} \braket{^+Ejj_3\eta}{\phi^+} 
\label{eq:lsket-expansion2}
\\
\Phi_+\ni\psi^- &= \sum_{j_3 \eta} \int_0^\infty dE \ket{Ejj_3\eta^-} \braket{^-Ejj_3\eta}{\psi^-}
\label{eq:lsket-expansion}
\end{align}
\label{eq:lsket-expansion-whole}
\end{subequations}
where $\braket{^\pm Ejj_3\eta}{\psi^\pm}$ fulfill 
Eq.\ \eqref{hardy-wave}.
	This means $\Phi_-$ is the (abstract) Hardy space whose wave functions are all smooth Hardy functions $\braket{^+E}{\phi^+}\equiv\braket{^+Ejj_3\eta}{\phi^+}$ analytic in the lower complex half-plane, and $\Phi_+$ is the (abstract) Hardy space whose wave functions $\braket{^-E}{\psi^-}\equiv\braket{^-Ejj_3\eta}{\psi^-}$ are analytic in the upper half-plane.
	Consequently $\braket{\psi^-}{E^-}=\overline{\braket{^-E}{\psi^-}}$ are analytic in the lower half-plane.

	Summarizing, there are two reasons that lead to the same conclusion: 
	First, in the discussions of the foundations of quantum mechanics one distinguishes between the two notions of state and observable, but in the conventional mathematical formulation 
\cite{ref:xx} 
one identifies the set of states $\{\phi^+\}$ with the set of observables $\{\psi^-\}$ as $\{\phi^+\}=\{\psi^-\}=\Phi$ ($=\mathcal{H}$ for the orthodox von Neumann axioms).
	Second, in the heuristic formulation of scattering theory one distinguishes (by the $\pm i \epsilon$ in energy) between the two Lippmann-Schwinger kets with in-coming and out-going boundary conditions.
	But in conventional scattering theory one treats the $\ket{E^+}=\ket{E+i\epsilon^+}$ and the $\ket{E^-}=\ket{E-i\epsilon^-}$ as if they were the same kind of Dirac kets, though the $\pm i \epsilon$ require different ways of analytic continuation.
	To overcome these two incongruities we make one new hypothesis that
\begin{subequations}
\begin{align}
&\quad\textrm{set of in-states}\,\,\{\phi^{+}\}=\Phi_{-}\subset\mathcal{H}\subset\Phi^{\times}_{-}\\
&\textrm{set of out-observables}\,\,\{\psi^{-}\}=\Phi_{+}\subset\mathcal{H}\subset\Phi^{\times}_{+},
\end{align}
\label{eq:newaxiom}
\end{subequations} 
where $\Phi_\mp$ are the two Hardy spaces of the semiplanes $\mathbb{C}_\mp$.
	This gives the Lippmann-Schwinger kets a precise mathematical meaning
\footnote{
The miss-match in the $\pm$ labels is due to the most conventional notations in physics for the in($+$) and out($-$) vectors, and in mathematics for the Hardy spaces. 
	Except for this miss-match in notation, the mathematics fits wonderfully for the physics; an example of what Wigner called the ``miracle of the appropriateness of the language of mathematics for the formulation of physics''.
}:  
\begin{eqnarray*}
\ket{E\mp i \epsilon^\mp} \equiv \ket{Ejj_3\eta^\mp} \in \Phi_\pm^\times.
\end{eqnarray*}
where $\Phi_\pm^\times$ are the duals of the Hardy spaces.
	The eigenket equation 
\eqref{eq:eigen-ls} 
(and similarly for 
\eqref{eq:diracket}) 
means mathematically precisely
\begin{align}
\braket{H\psi^-}{Ejj_3\eta^-}&\equiv \bra{\psi^-}H^\times\ket{Ejj_3\eta^-}
\nonumber
\\
&=E\braket{\psi^-}{Ejj_3\eta^-}
\label{eq:def-hconj}
\end{align}
for all $\psi^- \in \Phi_+$.
	The first $\equiv$ in 
Eq.\ \eqref{eq:def-hconj} 
uniquely defines the conjugate operator $H^\times$ as the extension to the space $\Phi_+^\times$ of the self-adjoint Hilbert space operator $H=H^\dagger \subset H^\times$.

	Since $\braket{\psi^-}{E^-}$ can be analytically continued into the lower complex semiplane and so can $\braket{H\psi^-}{E^-}$ (since also $H\psi^- \in \Phi_+$), one can continue 
Eq.\ \eqref{eq:def-hconj} 
into the lower semiplane to the value $z$ (unless $z$ is a singular point) and obtains
\begin{eqnarray}
\braket{H\psi^-}{zjj_3\eta^-} = z \braket{\psi^-}{zjj_3\eta^-}.
\label{eq:eigenvalue-z}
\end{eqnarray}
	The Gamow ket 
\eqref{eq:nonrel-gv} 
is not an analytic continuation of the Lippmann-Schwinger equation, but its singularity.

	If the Hamiltonian is explicitly known one can solve the time independent Schr\"{o}dinger equation 
\eqref{eq:eigenvalue} 
under the purely out-going boundary conditions and determine the solutions and their complex eigenvalues $z_{R_n} =E_{R_n}-i\Gamma_n/2$, see, e.g., for square well 
Refs.\ \onlinecite{ref:xxxx} and \onlinecite{ref:xxxxx}.
	They can be shown to coincide with the pole positions of the S-matrix
\cite{ref:xxxxx}.
	Alternatively, one can start from the pole of the S-matrix at $z_R=E_R-i\Gamma/2$ and obtain the Gamow vector 
\eqref{eq:nonrel-gv} 
as the pole term and then derive 
Eq.\ \eqref{eq:eigenvalue} 
from 
Eq.\ \eqref{eq:nonrel-gv} 
\cite{ref:28}. 
	The latter is what we shall do for the relativistic case in Sec.\,\ref{sec:4} (because in that case there is no Schr\"{o}dinger equation to solve).

	The time evolution of the Gamow vector $\psi^{G}(t)=e^{-iH^{\times}t}\psi^{G}$ can be derived 
\cite{ref:28} 
using the definition 
\eqref{eq:integral1}.
	It is given by
\begin{align}
\braket{e^{iHt}\psi^-_\eta}{\psi^G} 
&\equiv
\braket{e^{iHt}\psi_\eta^-}{z_R jj_3\eta^-} \notag \\
&\equiv
\bra{\psi_\eta^-} e^{-iH^\times t} \ket{z_Rjj_3\eta^-} \notag \\
&=
e^{-iE_R t}e^{-\Gamma/2 t}\braket{\psi_\eta ^-}{z_Rjj_3\eta^-},
\label{eq:timeevolution}
\end{align}
for all $\psi_{\eta}^{-} \in \Phi_{+}$, but for $t \geq 0$.
	For this derivation the hypothesis 
\eqref{eq:newaxiom} 
is essential
\footnote{
The time asymmetry $t\geq 0$ has its mathematical origin in the Paley-Winter theorem for Hardy spaces, whereas the reversible unitary group evolution $-\infty <0<+\infty$ for the Hilbert space follows from the Stone-von Neumann theorem.
	For more on time asymmetry see, e.g.,  
Refs.\ \onlinecite{ref:B3,ref:B52,ref:B53}.
}.
	Because of the properties of the   
Eq.\ \eqref{eq:timeevolution},  
we call the ket $\psi^G$ a Gamow ket.
	It has the properties envisioned by Gamow, namely exponential time evolution.
	From 
Eq.\ \eqref{eq:timeevolution} 
one sees that $\psi^{G}$ is the exponentially decaying state with the lifetime $\tau=1/\Gamma$, where $\Gamma$ is the Breit-Wigner width Breit-Wigner in 
Eqs.\ \eqref{eq:integral1} and \eqref{eq:nonrel-gv}, and the width of the line shape $|a_{j}^{BW}(E)|^{2}$ of 
\eqref{eq:bwamp}. 
	The Gamow ket $\psi^{G} \in \Phi_{+}^{\times}$ has all the properties one wanted in a state vector for quasistable particle; its energy distribution has the width $\Gamma$ and its lifetime is $\hbar/\Gamma$.
	The Gamow ket 
\eqref{eq:nonrel-gv} 
thus unifies resonance scattering and decay in non-relativistic quantum mechanics.

\section{\label{sec:4}Relativistic Resonances}

\subsection{\label{subsec:4-1}The relativistic in- and out- Lippmann-Schwinger kets and the S-matrix}

	In order to obtain a unique definition of a relativistic resonance and to combine it with the notion of a decaying state, we have to define a relativistic Gamow vector.
 	For this we combine the results of Section \ref{sec:2} with the concepts of Section \ref{sec:3}.
	In Section \ref{sec:2} the relativistic Breit-Wigner amplitude 
\eqref{eq:relbwamp} 
emerged as the favored resonance amplitude.
	This means that the relativistic resonance is defined by a (first order
\footnote{
Higher order poles can also be included 
\cite{ref:new33a}
but for the sake of simplicity we restrict ourselves here to first order resonances.}) 
pole of the $j$-th partial S-matrix $S_j^{n'n}(s)$ where $j$ is the spin of the resonance, and we will associate to each pole of the S-matrix $s_{R_i}$ a relativistic Gamow vector in very much the same way as it was done in 
Eq.\ \eqref{eq:nonrel-gv}.
	For this we need the relativistic Lippmann-Schwinger kets of scattering processes.

	In order to include in our discussion the superposition of resonances and the interference between decaying states, we consider scattering experiment with two resonances in the $j$-th partial wave.
	The generalization to a finite (or even infinite \cite{ref:gad}) number of resonances is straight forward.
 
	According to the phenomenological results in Section \ref{sec:2} there is always a background $B_j$.
	This means the scattering goes through two resonances $R_1$ and $R_2$ and the background $B$ (e.g., the direct production of 
Eq.\ \eqref{nonresonant}):
\begin{eqnarray}
1+2 
\longrightarrow 
\begin{Bmatrix} R_1 \\ R_2 \\ B \end{Bmatrix}
\longrightarrow 
3+4.
\label{eq:formation}
\end{eqnarray}
For instance, $(1,2)=(e^+, e^-)=n$ and $(3,4)=(f, \bar{f})=n'$, where $n$ and $n'$ denote particle species quantum number.
	The accelerator prepares a two-particle in-state $\phi^{in}$ and the detector registers the two out-particles $\psi^{out}$:
\begin{subequations}
\begin{align}
\phi^{in}\,\,=\ket{e^+ e^-} &\longrightarrow \phi^+_n \,=\Omega^+ \phi^{in},
\\ 
\psi^{out}=\ket{f \bar{f}}\quad &\longleftarrow \psi^-_{n'}=\Omega^- \psi^{out}. 
\end{align}
\label{ee-ff}
\end{subequations}
	The matrix element $(\psi^-_{n'},\phi^+_{n})$ is the Born probability amplitude for the out-observable $\psi^-$ in the prepared in-state $\phi^+$.
	It is usually written as the S-matrix element $(\psi_{n'}^-, \phi_n^+)=(\psi^{out},S\phi^{in})$.
	For the in-state $\phi^+$ and the out-observable  $\psi^-$, we use the new Hardy space hypothesis 
\eqref{eq:newaxiom} 
with the energy $E$ now replaced by the relativistic variable $s=(p_1+p_2)^\mu (p_1+p_2)_\mu = p^\mu p_\mu$.

	The prepared in-state $\phi^+ \in \Phi_-$ and the registered out-observables $\psi^- \in \Phi_+$ are expanded with respect to the basis systems
\footnote{
Here $[s,j]$ labels the irreducible representations of Poincar\'{e} transformation.
Instead of the total momentum $p=(p^0,\mathbf{p})=p_1+p_2=p_3+p_4$, we choose to label the basis kets of the representation $[s,j]$ by the three space components of the four-velocity \cite{ref:33} 
$\vel{p} \equiv
\frac{\mathbf{p}}{\sqrt{s}}
=\gamma \mathbf{v}
=\frac{1}{\sqrt{1-\mathbf{v}^2}} \mathbf{v}$ and 
$\hat{p}_0(\vel{p})\equiv
\frac{p^0}{\sqrt{s}}=
\gamma$, where $\mathbf{v}$ denotes the three velocity.
	The invariant integration measure $\velm{p}$ in 
Eqs.\ \eqref{eq:lsket-expansion-phi} and \eqref{eq:lsket-expansion-psi}  
had to be chosen correspondingly, and there is no difference between using the $\vel{p}$ as labels of the kets or using the momentum $\mathbf{p}$ as the degeneracy labels in an irreducible representation $[s,j]$.
}
\begin{subequations}
\begin{align}
\phi_n^+ &=\int_{(m_1+m_2)^2}^{\infty} ds 
\notag
\\
&\quad\times \sum_{jj_3} \int_{-\infty}^{+\infty} \velm{p} 
\ket{[s,j]\mathbf{\hat{p}}j_3n^+}
\braket{^+nj_3\mathbf{\hat{p}}[s,j]}{\phi^+} \notag \\
&\equiv\int_{s_0}^\infty ds \ket{s^+}\braket{^+s}{\phi^+}, 
\label{eq:lsket-expansion-phi}
\\
\psi_n^- &= \int_{(m_3+m_4)^2}^{\infty} ds
\notag
\\
&\quad \times \sum_{jj_3} \int_{-\infty}^{+\infty} \velm{p}
\ket{[s,j]\mathbf{\hat{p}}j_3n^-}
\braket{^-nj_3\mathbf{\hat{p}}[s,j]}{\psi^-} \notag \\
&\equiv\int_{s_0}^\infty ds \ket{s^-}\braket{^-s}{\psi^-}.
\label{eq:lsket-expansion-psi}
\end{align} 
\label{eq:lsket-expansion-whole-rel}
\end{subequations}
%
	This is the relativistic analogue of the non-relativistic basis vector expansion in 
Eqs.\ \eqref{eq:lsket-expansion2} 
and 
\eqref{eq:lsket-expansion}.
	The new Hardy space axiom 
\eqref{eq:newaxiom} 
in the relativistic case means that the relativistic energy wave functions (as functions of $s$)
\footnote{
With the choice of the integration in 
Eq.\ \eqref{eq:lsket-expansion-psi} 
the ``normalization''
of the basis kets is $\braket{^-\vel{p}'j'_3n'[s',j']\,}{\,\vel{p}j_3n[s,j]^-}
=2\hat{p}^0 (\vel{p})\delta^3(\vel{p}'-\vel{p})
\delta(s'-s)\delta_{j'_3 j_3}\delta_{j'j}$.
	When we make the analytic continuation in $s$ the momenta also become complex.
	But we choose only those complex mass representations of Poincar\'{e} transform for which the four-velocity $\vel{p}=\frac{\mathbf{p}}{\sqrt{s}}$ remains real (``minimally complex representations '') which agree with those of 
Ref.\ \onlinecite{ref:35}.
},
\begin{subequations}
\begin{align}
&\braket{^+s}{\phi^+}\equiv\braket{^+ n j_3 \mathbf{\hat{p}}[s,j]}{\phi^+}\,=\overline{\braket{^+\phi}{s^+}},
\label{eq:hardy-phi}
\\
&\braket{^-s}{\psi^-}\equiv\braket{^-nj_3\mathbf{\hat{p}}[s,j]}{\psi^-}=\overline{\braket{^-\psi}{s^-}},
\label{eq:hardy-psi}
\end{align}
\label{eq:hardy-whole}
\end{subequations}
are Hardy functions. 
	Specifically, the wave functions in 	
Eq.\ \eqref{eq:hardy-phi} 
are analytic in the lower complex $s$-plane (second Riemann sheet of the S-matrix) and those in 
Eq.\ \eqref{eq:hardy-psi} 
are analytic in the upper $s$-plane, i.e., $\braket{^-\psi}{s^-}$ are analytic in the lower half plane.

	The relativistic Lippmann-Schwinger kets $\ket{[s,j] \hat{\mathbf{p}} j_{3} \eta^\pm}$ are very similar to the basis vectors obtained in the direct product of the two Poincar\'{e} group representations $[m_1,j^{(1)}]\times[m_2, j^{(2)}]$ (or $[m_3,j^{(3)}]\times[m_4, j^{(4)}]$) used in the relativistic partial wave expansion 
\cite{ref:32,ref:33}, 
however, here they are not ordinary Dirac kets $\ket{[s,j]\mathbf{p}j_3n}$ but elements of the spaces $\Phi_\mp^\times$.
	The Poincar\'{e} generators (the momentum operators and the Lorentz generators) are ``the exact generators'' which include interactions 
\cite{ref:33a}.
	In place of the usual $\pm i \epsilon$ of quantum field theory, which is the imaginary part of energy $p^0=p_1^0+p_2^0$, our Lippmann-Schwinger kets have the $\pm i \epsilon$ as an addition to the invariant energy squared: $\ket{[s\pm i\epsilon,j]\mathbf{\hat{p}}j_3n^\pm}$. 
	As long as $i \epsilon$ in $s\pm i \epsilon$ is infinitesimal it makes no difference whether one uses $s\pm i \epsilon$ or $p^0\pm i \epsilon$ 
\footnote{
$s=(p^0\pm i \epsilon')^2-\mathbf{p}^2=s\pm i 2p^0 \epsilon'-\mathbf{p}^2=s\pm i \epsilon -\mathbf{p}^2$.
}, 
but when we analytically continue to values of $s$ in the whole complex semiplane, we want to use a Lorentz invariant complex variable $s$. 
	Also, in place of the momentum $\mathbf{p}$ we use the dimensionless $\mathbf{\hat{p}}=\frac{\mathbf{p}}{\sqrt{s}}$ to label the kets in an irreducible representation $[s,j]$.

	The basis vectors
\begin{eqnarray}
\ket{s^\pm} \equiv \ket{[s,j]\mathbf{\hat{p}}j_3n^\pm}
\in \Phi_\mp^\times
\label{eq:rel-basis-vect}
\end{eqnarray}
in 
Eqs.\ \eqref{eq:lsket-expansion-phi} and \eqref{eq:lsket-expansion-psi} 
span the direct product space of two out going ($-$) and incoming ($+$) particles.
	The possible physical values of $[s,j]$ are $(m_1+m_2)^2 \leq s < \infty$ and $j=j^1+ j^2, j^1+ j^2 +1, j^1+ j^2 +2, \ldots$, where $m_1,m_2$  are the masses and $j^1,j^2$ are the spins of the incoming particles 
\cite{ref:32,ref:33}.
	Like the basis vectors $\ket{[m^2,j]\mathbf{\hat{p}}j_3n}$ of an irreducible unitary representation $[m^2,j]$ of the Poincar\'{e} group
\cite{ref:30a} 
the vectors 
\eqref{eq:rel-basis-vect}
also transform irreducibly (keeping the value $[s,j]$ unchanged) under Poincar\'{e} transformations.
	But they do {\it not} furnish unitary group representations (see below).

	If one inserts 
Eqs.\ \eqref{eq:lsket-expansion-phi} and \eqref{eq:lsket-expansion-psi} 
into the S-matrix element $(\psi^-_{n'},\phi^+_n)$ and uses invariance of the S-matrix with respect to Poincar\'{e} transformations, one obtains the Born probability amplitude in terms of the S-matrix elements $S^{n'n}_j(s)$ with angular momentum $j$:
\begin{align}
(\psi_{n'}^-, \phi_n^+)
&=
\int_{m_0^2}^\infty ds 
\sum_{jj_3} \int_{-\infty}^{+\infty} \velm{p}
\nonumber
\\
&\,
\times \braket{\psi^-}{[s,j]j_3\mathbf{\hat{p}}{n'} ^-} S_j^{n'n}(s) \braket{^+ n j_3 \mathbf{\hat{p}}[s,j]}{\phi^+}. \nonumber\\
\,
\label{eq:rel-s-matrix-element}
\end{align}
	This $j$-th partial S-matrix element $S_j^{n'n}(s)$ is the reduced matrix element of the S-matrix  defined by
\begin{eqnarray}
&&\braket{^-\vel{p}'j'_3n'[s',j']\,}{\,\vel{p}j_3n[s,j]^+}
\nonumber
\\
&&\equiv
\bra{\,\vel{p}'j'_3n'[s',j']\,} \, S \, \ket{\,\vel{p}j_3n[s,j]\,}
\notag \\
&&=
2\hat{p}^0 \delta^3(\vel{p}'-\vel{p})
\delta(s'-s)\delta_{j'_3 j_3}\delta_{j'j}S^{n'n}_j(s),
\nonumber
\\
\label{eq:s-matrix-element}
\end{eqnarray}
after the Poincar\'{e} invariance has been taken into account and expressed in terms of $\delta$-function for the continuous label and the Kronecker-$\delta$ for the discrete one.
	For the ``continuous summation'' we used the Lorentz invariant measure $\velm{p}$ of the Dirac basis system (Nuclear Spectral theorem) \eqref{eq:lsket-expansion-phi} and \eqref{eq:lsket-expansion-psi}.
	To prove Eq.\ \eqref{eq:s-matrix-element} one does not need the whole Poincar\'{e} group.

\subsection{\label{subsec:4-2}The property of the $j$-th partial S-matrix $S_j(s)$}

	After the Poincar\'{e} invariance has been taken into account the Poincar\'{e} labels $j_3$ and $\vel{p}$ are of no further importance.
	Therefore, for the far r.h.s of 
Eqs.\ \eqref{eq:lsket-expansion-phi} and \eqref{eq:lsket-expansion-psi} 
we have used a truncated notation and suppressed the labels $j_3$ and  $\vel{p}$  that label the basis vectors within an irreducible Poincar\'{e} representation space $[s,j]$.
	We also suppressed the angular momentum $j$ and the species quantum numbers $n$ because we shall restrict ourselves to the partial wave with fixed resonance spin $j$.
	In this abbreviated notation we write the $j$-th term in the sum in 
Eq.\ \eqref{eq:rel-s-matrix-element} 
as
\begin{eqnarray}
(\psi^-,\phi^+)_j 
=
\int_{m_0^2}^\infty ds \braket{\psi^-}{s^-}S_j(s) \braket{^+s}{\phi^+}.
\label{eq:rel-s-matrix-element-simple}
\end{eqnarray}
The $j$-th partial S-matrix element $S_j(s)$ describes the dynamics of the scattering process.
	We assume for $S_j(s)$ the standard analyticity properties of the S-matrix.
	It is connected with the scattering amplitude of 
Eq.\ \eqref{eq:amp}:
\begin{subequations}
\begin{align}
S_j(s)&=2ia_j(s)+1 \quad \mbox{for elastic channels},
\\
S_j(s)&=2ia_j(s) \quad \quad \,\,\,\,\mbox{for inelastic channels}.
\end{align}
\label{eq:s-matrix-amp}
\end{subequations}
	A resonance has a definite spin, $j_R$, phenomenologically, i.e., resonances appear in a particular partial wave $j=j_R$, which is the one we have selected in 
Eq.\ \eqref{eq:rel-s-matrix-element-simple}.

	The same resonance can appear in different channels, but we have chosen in 
Eq.\ \eqref{eq:rel-s-matrix-element-simple} 
one particular channel $n'$ by fixing the particle species labels $(n',n)$ in 
Eq.\ \eqref{eq:rel-s-matrix-element}. 
	When we burrow down through the cut along the real axis, we will then be on one particular Riemann sheet above the $n'$-threshold, the ``unphysical'' sheet, in which the resonance pole $s_{R_1}$ and $s_{R_2}$ are located. 
	This we called the second Riemann sheet, but it could also be one of the higher ``unphysical'' sheets.

	There can be more than one resonance (more than one pole) in the same partial wave.
	We are considering in 
Eq.\ \eqref{eq:formation} 
the case of two resonance poles located at different positions $s=s_{R_1}$ and $s_{R_2}$.
	For the sake of simplicity we assume that there are only the two first order poles on the second sheet of $S_j(s)$ and we assume that the two poles at $s=s_{R_1}$ and at $s=s_{R_2}$ are sufficiently close to each other and to the cut along the real axis from $m_{n'}^2=m_0^2=(m_3+m_4)^2 \leq s < \infty$.
	This is the situation depicted in Fig.\ 1.

	The integration in Eq.\ \eqref{eq:rel-s-matrix-element-simple} is done along the lower edge of the first sheet which is the same as the upper edge of the second sheet.
	The Hardy property postulated by the new Hardy space axiom \eqref{eq:newaxiom} refers to the analyticity property on the second (or higher) Riemann sheet of $S_j(s)$.
	This means that according to the new axiom \eqref{eq:newaxiom} the energy wave functions of the prepared in-state $\phi^+(s)$ and the complex conjugate of the detected out-observable $\overline{\psi^-}(s)$:
\begin{subequations}
\begin{align}
\phi^+(s) &\equiv \braket{^+s}{\phi^+},
\label{eq:rel-newaxiom-phi}
\\
\overline{\psi^-}(s) &\equiv \braket{\psi^-}{s^-}=\overline{\braket{^-s}{\psi^-}}
\label{eq:rel-newaxiom-psi}
\end{align}
\label{eq:rel-newaxiom}
\end{subequations}
are smooth Hardy functions on the lower complex $s$ plane (second sheet).
	This axiom and the properties of Hardy function are essentially all that we need for the following derivations.

	Without this axiom 
\eqref{eq:newaxiom} 
we cannot derive the superposition of two Breit-Wigner amplitudes or of two Gamow vectors for which there is sufficient experimental evidence, e.g., for the $Be^8$ nucleus at 16.6$MeV$ and 16.9$MeV$ 
\cite{ref:B78} 
or for the neutral Kaon system
\cite{ref:B19}.
	The same Hardy space axiom 
\eqref{eq:newaxiom} 
is also required to derive the Gamow vector from the pole term.
	Except for this new axiom, all other assumptions which we shall use are the standard axioms of quantum theory and relativistic invariance.
	In particular for $S_j(s)$ we shall make the standard assumption of polynomial boundedness and analyticity 
\cite{ref:10}.


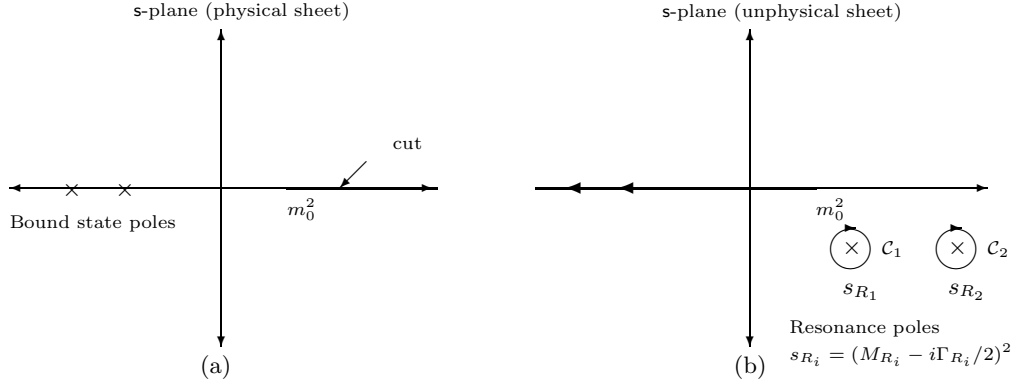
\begin{figure*}

\begin{picture}(400, 150)(0,0)

\put(100,75){\vector(0,1){60}}
\put(100,75){\vector(0,-1){60}}
{\thicklines \put(125,75){\line(1,0){57}}}
\put(100,75){\vector(1,0){80}}
\put(100,75){\vector(-1,0){80}}

\put(67, 140){\scriptsize $\mathsf s$-plane (physical sheet)}
\put(125, 65){\scriptsize $m_0^2$}
\put(89, 5){ (a)}
\put(294, 5){(b)}

\put(60, 72){\small $\times$}
\put(40, 72){\small $\times$}

\put(20, 60){\scriptsize Bound state poles}

\put(155,85){\vector(-1,-1){10}}
\put(165,90){\scriptsize cut}

\put(300,75){\vector(0,1){60}}
\put(300,75){\vector(0,-1){60}}
{\thicklines \put(325,75){\line(-1,0){106}}}
\put(300,75){\vector(1,0){90}}
\put(300,75){\line(-1,0){80}}
{\thicklines \put(240,75){\vector(-1,0){10}}}
{\thicklines \put(260,75){\vector(-1,0){10}}}

\put(267, 140){\scriptsize $\mathsf s$-plane (unphysical sheet)}
\put(325,65){\scriptsize $m_0^2$}

\put(335, 50){\small $\times$}
\put(335, 35){$ s_{R_1}$}
\put(338, 52){\circle{15}}
\put(350, 50){\scriptsize $ \mathcal{C}_1$}
\put(337, 60){\vector(1,0){3}}

\put(375, 50){\small $\times$}
\put(375, 35){$ s_{R_2}$}
\put(378, 52){\circle{15}}
\put(390, 50){\scriptsize $ \mathcal{C}_2$}
\put(377, 60){\vector(1,0){3}}

\put(315, 20){\scriptsize Resonance poles}
\put(315, 10){\scriptsize $s_{R_i} = (M_{R_i} -i\Gamma_{R_i}/2)^2$}

\end{picture}

\caption{
	The two sheeted S-matrix. The $j^{th}$ partial S-matrix
$S_j(s)$ is an analytic function on a Riemann energy surface cut along
the positive real axis from $m_0^2 \leq s < \infty$ indicated
in Fig.\ 1a.
	The integration in 
Eq.\ \eqref{eq:rel-s-matrix-element} 
is along the cut in Fig.\ 1a, either on the
lower edge of the ``physical sheet'' or along the upper edge of the
second sheet.
	The contour of integration can be deformed into the lower half plane of the second sheet, and ultimately into the contours around the two resonance poles indicated by $\times$ and into an integral from $m_0^2$ to $-\infty_{II}$ along the upper edge of the second sheet. 
	This is shown in Fig.\ 1b; the arrows indicate the direction of integration.
Thus we have the equality of the integrals 
Eqs.\ \eqref{eq:rel-s-matrix-element} and \eqref{eq:cont-deform}.
	For the non-relativistic case the picture is similar, except that one has to identify $E=E_0=0$ with $s=m_0^2$.}
\end{figure*}

\subsection{\label{subsec:4-3}The relativistic Gamow kets associated to the resonance pole}

	We shall derive now the properties of relativistic resonances and decaying states from the pole definition.
	The two resonances are introduced 
by a first order pole  at the position ${s}={s}_{R_1}$ and ${s}={s}_{R_2}$ in the 
second sheet. 
	As a consequence of the Hardy space assumption, specifically of Eqs.\ \eqref{eq:hardy-phi} and \eqref{eq:hardy-psi}, the integrand in 
Eq.\ \eqref{eq:rel-s-matrix-element-simple} 
is analytic in the lower half plane of the second sheet except for the two poles at ${s}={s}_{R_i}$. 
	The contour of integration of 
Eq.\ \eqref{eq:rel-s-matrix-element-simple} 
is depicted in
Fig.\ 1a; 
it is the cut along the real axis from $(m_1+m_2)^2 \equiv m_0^2 \leq s < \infty$. 
	We deform the contour of  integration in 
Eq.\ \eqref{eq:rel-s-matrix-element-simple} 
from the positive real line on the first sheet through the cut into the lower half plane of the second sheet. 
	The integral over the infinite semicircle is omitted since it is zero as a consequence of the Hardy space hypothesis 
\eqref{eq:newaxiom} 
and boundedness property of $S_j(s)$.  
	The result of this contour deformation is shown in 
Fig.\ 1b 
and we obtain for 
Eq.\ \eqref{eq:rel-s-matrix-element-simple}:
\begin{align}
(\psi ^-,\phi ^+)
&=\int_{m_0^2}^{-\infty _{II}}d{s}\,
  \braket{\psi ^-}{s^-} S_{II}(s)
  \braket{^+ s}{\phi ^+}  
\nonumber
\\
&+\oint_{C_1} ds\, \braket{\psi ^-}{s^-} S_{II}({s})
   \braket{^+s}{\phi ^+}
\nonumber 
\\
&+\oint_{C_2} d{s} \, 
  \braket{\psi ^-}{s^-} S_{II}({s})
  \braket{^+s}{\phi ^+}.
\label{eq:cont-deform}
\end{align}
	The kets $\ket{s^-}$ and the bras $\bra{^+s}$ are the analytic continuation of the Lippmann-Schwinger kets $\ket{s_{real}-i\epsilon}$ and the Lippmann-Schwinger bras $\bra{s_{real}+i\epsilon}$ into the complex $s$-plane second sheet of the S-matrix element $S_j^{n'n}(s)$, except for the singular points $s_{R_i}$.
	The ket $\ket{s^-}$ can be continued into the lower half-plane (where $\braket{\psi^-}{s^-}$ is analytic) and the bra $\bra{^+s}$ can also be continued into the lower half-plane (where $\braket{^+s}{\phi^+}$ is analytic). 
	We have omitted the subscript $j$ at $(\psi^-,\phi^+)_j$ and at $S_{II_j}(s)$, the subscript $II$, again, means we are now on the 2nd sheet.
 	$C_i$ denotes the circle around the pole at ${s}_{R_i}$, and the first integral extends along the negative real axis in the second sheet (indicated by $-\infty _{II}$).
	The first term has nothing to do with any of the resonances, it is the non-resonant background term, 
\begin{equation}
       \int_{m_0^2}^{-\infty _{II}}d{s}\, 
       \langle \psi ^-|{s}^-\rangle S_{II}({s})
       \langle ^+{s}|\phi ^+ \rangle \equiv 
       \langle \psi ^-|\phi ^{bg}\rangle \, .
\label{eq:background}
\end{equation}
which we express as the matrix element of $\psi^-$ with a generalized vector $\phi^{bg}$ that is defined by 
Eq.\ \eqref{eq:background} 
(as a functional on $\Phi_+=\{\psi^-\}$):
\begin{equation}
\phi^{bg} \equiv \int_{m_0^2}^{-\infty} ds \kt{s^-} \bk{^+s}{\phi^+} S_{II}(s).
\label{eq:background-vect}
\end{equation}
	We will return to it below.
	We now consider each integral along $C_i$ around each pole at $s_{R_i}$ separately. 
	For each integral separately we use the expansion around the pole ${s}_{R_i}$:
\begin{equation}
      S(s)=\frac{R^{(i)}}{{s}-{s}_{R_i}}+R_0+
      R_1({s}-{s}_{R_i})+\cdots \,.
      \label{37}
\end{equation}
	For each of the two (or $N$) integrals {\it separately} we evaluate the integrals around each pole  $s_{R_i}$. 
	Then we obtain for each of these pole terms the following results:
\begin{eqnarray}
      (\psi ^-,\phi ^+)_{\rm pole_i} &=&
      \oint_{\hookleftarrow C_i} d{s}\, 
      \langle \psi ^-|{s}^-\rangle S({s})
       \langle ^+{s}|\phi ^+ \rangle \nonumber \\
      &=&\oint_{\hookleftarrow C_i} d{s}\, 
      \langle \psi ^-|{s}^-\rangle 
      \frac{R^{(i)}}{{s}-{s}_{R_i}} 
       \langle ^+{s}|\phi ^+ \rangle \nonumber \\ 
      &=&-2\pi i R^{(i)} \langle \psi ^-|{s}_{R_i}^-\rangle 
          \langle ^+{s}_{R_i}|\phi ^+ \rangle  
\label{eq:pole-1}
\\
      &=& \!\!\! \int_{-\infty _{II}}^{\infty}\! \!\!\!\! d{s} \,
       \langle \psi ^-|{s}^-\rangle 
       \langle ^+{s}|\phi ^+ \rangle 
       \frac{R^{(i)}}{{s}-{s}_{R_i}} \, .
\label{eq:pole-2}  
\end{eqnarray}
	This simple derivation is possible only if one makes use of the Hardy property of the wave functions and uses the theorems of Cauchy (for 
Eq.\ \eqref{eq:pole-1}) 
and of Titchmarch (for 
Eq.\ \eqref{eq:pole-2}).
	Inserting 
Eqs. \eqref{eq:background} and \eqref{eq:pole-1} 
into  
Eq.\ \eqref{eq:cont-deform} gives the following representation of the $S$-matrix element 
\eqref{eq:cont-deform}:
\begin{equation}
       (\psi ^-,\phi ^+)=\langle \psi ^-|\phi ^{bg}\rangle +
       \sum_i \langle \psi ^-|{s}_{R_i}^-\rangle \frac{2\pi R^{(i)}}{i}
       \langle ^+{s}_{R_i}|\phi ^+\rangle \, .
\label{eq:bg-and-pole}
\end{equation}
	This has introduced a new ket $\ket{s_{R_i}^-}$ for every S-matrix pole at the singularity $s=s_{R_i}$.
	The vector $\psi^- \in \Phi_+$ represents the out-particles observed by the detector, e.g., the decay products $\mu^+\mu^-$ of the $Z^0$ resonance in the resonance scattering process 
\eqref{eq:formation}.
	We can omit $\psi^- \in \Phi_+$ from 
Eq.\ \eqref{eq:bg-and-pole} 
and obtain the same statement 
Eq.\ \eqref{eq:bg-and-pole} 
as an equation between generalized vectors in the space $\Phi_+^\times$.
	This gives a new basis vector expansion of the prepared in-state
$\phi^+ \in \Phi_-$ in terms of eigenvectors with complex eigenvalues:
\begin{eqnarray}
       \phi ^+&=&\phi ^{bg}+\sum_i|{s}_{R_i}^- \rangle c_{R_i}.
\label{eq:bg-and-pole-simple}
\end{eqnarray}
	Here the expansion coefficients are given by
\begin{eqnarray*}
c_{R_i}=  (2\pi R^{(i)}/i) \langle ^+{s}_{R_i}|\phi ^+\rangle \, ,
\end{eqnarray*}
and $\phi^{bg}$ is given by the integral 
\eqref{eq:background-vect}.
	This {\it complex basis vector expansion} is an alternate to the basis
vector expansion in 
Eq.\ \eqref{eq:lsket-expansion-phi}.

	We first consider one of the vectors $\kt{s_{R_i}^-}$ in the discrete sum in 
Eq.\ \eqref{eq:bg-and-pole-simple}.
	From the equality 
Eq.\ \eqref{eq:pole-1}=Eq.\ \eqref{eq:pole-2} 
we see that one can define a whole class of generalized vectors the $\kt{s_{R_i}^-}$:
\begin{eqnarray}
\braket{\psi^-}{s_{R_i}^-}_{\phi^+}
\equiv
\frac{i}{2\pi} \int_{-\infty_{II}}^{+\infty}\!\!\! ds \braket{\psi^-}{s^-}
\frac{\braket{^+s}{\phi^+}}{\braket{^+s_{R_i}}{\phi^+}}
\frac{1}{s-s_{R_i}}
\nonumber
\\
\label{eq:rel-gv-func}
\end{eqnarray}
as functionals of $\psi^- \in \Phi_+$.
	Of these generalized vectors we single out one vector with a particular ``normalization'': 
\begin{eqnarray}
\ket{s_{R_i}^-} 
&\equiv& 
\frac{i}{2\pi}\int_{-\infty_{II}}^{+\infty} ds \ket{s^-}\frac{1}{s-s_{R_i}}.
\label{eq:rel-gamow-vect-simple}
\end{eqnarray}
	This ket (functional on the Hardy space $\Phi_+=\{\psi^-\}$) we call the relativistic Gamow vector.
	Returning to complete labels of the basis vectors as in 
Eq.\ \eqref{eq:lsket-expansion-phi} 
this definition is written as
        \begin{equation}
                \kt{s_{R_i}^-}= \kt{[s_R,j]\mathbf{\hat{p}} j_3 ^-} 
                \equiv \frac{i}{2\pi} \int_{-\infty_{II}}^{+\infty} ds
                \kt{[s,j] \mathbf{\hat{p}}j_3 ^-} \frac{1}{s-s_R}.
\label{eq:rel-gamow-vect}
        \end{equation}
	Its normalization follows from that of the Lippmann-Schwinger kets, which is connected to the choice of the integration measure in 
Eq.\ \eqref{eq:lsket-expansion-psi}.

        In contrast to the integration boundaries  $m_0^2 \leq s < +\infty$  in the basis vector expansion 
\eqref{eq:lsket-expansion-psi} 
for the proper vectors $\psi_j^- \in \Phi_+$, the integration in 
Eq.\ \eqref{eq:rel-gamow-vect} extends from $-\infty_{II} <s < +\infty$, i.e., 
it extends over the real energy axis on the second sheet, which coincides  for $s \geq m_0^2$ with the physical values on the lower edge of the first sheet, see  
Fig. 1.
	As in the non-relativistic case \eqref{eq:nonrel-gv}, this indicates that the generalized vectors $\kt{[s_R,j]\mathbf{\hat{p}} j_3 ^-}\equiv \ket{s_R^-}$ are Hardy space functionals, i.e., elements of the dual space $\Phi_+^\times$. 
        The value $s_R$ in 
Eq.\ \eqref{eq:rel-gamow-vect} and in Eq.\ \eqref{eq:rel-gamow-vect-simple} 
is the position of the resonance pole in 
Eq.\ \eqref{37} 
of the analytically continued S-matrix (which is on the second sheet of the Riemann surface).

	So far we have discussed resonance formations \eqref{resonances}.
	We have defined the Gamow vector for resonance formation only and derived the integral representation of the Gamow kets \eqref{eq:rel-gamow-vect} from the S-matirx poles for resonance formations \eqref{eq:formation} and \eqref{resonances}.
	Breit-Wigner bumps are also (and predominantly) observed in resonance production like
\begin{align}
a+b \longrightarrow c+R \longrightarrow c+e+f.
\label{eq:production2}
\end{align}
If the Gamow vector is the representation of the resonance $R$ per se using the Gamow vector \eqref{eq:rel-gamow-vect} for $R$ in the process \eqref{eq:production2} should lead to a Breit-Wigner factor in the amplitude of the process \eqref{eq:production2}, and a Breit-Wigner line shape factor in its modulus square.
	It can indeed be shown \cite{ref:production1} that this is the case and that the amplitude for the process \eqref{eq:production2} contains a Breit-Wigner amplitude in the invariant energy square of the two-particle system $e+f$,
\begin{align}
s_{ef}=(p_e+p_f)^2=(p_{ab}-p_c)^2=(p_a+p_b-p_c)^2,
\end{align}
given by
\begin{align}
\frac{1}{s_{ef} - s^*_R} = \frac{1}{(p_{ab}-p_c)^2 - s^*_R}
\end{align}
where $s_R=(M_R-i\Gamma_R/2)^2$ is the complex mass of the Gamow state vector defined by Eq.\ \eqref{eq:rel-gamow-vect}.
	This shows that resonance formation and resonance production have their origin in the same physical entity described by the semigroup representation $[s_R,j]$ of the Poincar\'{e} transformations and related to the S-matrix pole $s_R$ by Eq.\ \eqref{eq:rel-gamow-vect} \cite{ref:production2}.

\subsection{\label{subsec:4-4}Properties of the relativistic Gamow vector under Poincar\'{e} transformations}

        The vectors \eqref{eq:rel-gamow-vect} 
which emerged from the resonance pole at $s_{R_i}$, are defined in complete
        analogy to the non-relativistic Gamow vectors 
\eqref{eq:nonrel-gv}
        except that in place of $E$, the relativistically invariant energy
        square $s$ has been used. In the same way as for the non-relativistic
        case in 
Eq.\ \eqref{eq:eigenvalue}, 
one can show that the $\kt{[s_R,j] \vel{p} j_3\,^-}$ are generalized eigenvectors of the total
        invariant mass square operator $P_\mu P^\mu$ with complex eigenvalue $s_R$
\footnote{
$(P^\mu P_\mu)^\times \supset (P^\mu P_\mu)^\dagger$ is the uniquely defined extension of the self adjoint operator $(P^\mu P_\mu)^\dagger$ in $\mathcal{H}$ to the conjugate space $\Phi_+^\times \supset \mathcal{H}$.
}:
        \begin{equation}
                (P_\mu P^\mu)^\times \kt{[s_R,j] \vel{p}
        j_3\,^-} = s_R \kt{[s_R,j] \vel{p} j_3\,^-}.
\label{rel-ev-eq}
        \end{equation}
        Thus one has an association between the ``exact''
        relativistic Breit-Wigner amplitude \eqref{eq:relbwamp} 
extended to $s=-\infty_{II}$, and the space of
        relativistic Gamow 
        vectors, i.e., the space spanned by the vectors 
\eqref{eq:rel-gamow-vect} 
with a fixed value of $[s_R,j]$: 
\begin{eqnarray}
\label{association}
a_j^{BW}(s) = \frac{r}{s-s_R} \Longleftrightarrow \{\psi_{[s_R,j]}^G\}
\end{eqnarray}
where
\begin{eqnarray}
\psi_{[s_R,j]}^G 
&\equiv&
\sum_{j_3} \int_{-\infty}^{+\infty} \velm{p} \kt{[s_R,j] \vel{p} j_3\,^-} \psi_{j_3}(\vel{p})
\label{eq:rel-gamow-state}
\end{eqnarray}
for all $\psi_{j_3}(\mathbf{\hat{p}}) \in \mathcal{S}(\mathbb{R}^3)$ and $-j  \leq  j_3 \leq j$.
	Here $\mathcal{S}(\mathbb{R}^3)$ denotes the set of all smooth, rapidly decreasing functions of $\vel{p}$ (Schwartz space).
	The same kind of spaces can be formed with the Lippmann-Schwinger kets:
\begin{eqnarray}
\psi_{[s-i\epsilon,j]}^-
=
\sum_{j_3} \int_{-\infty}^{+\infty}  \velm{p} \ket{[s,j]\vel{p}j_3\,^-}\psi_{j_3}(\vel{p}).
\label{eq:psi-lsket}
\end{eqnarray}        
	We denote these spaces of generalized eigenvectors of $P^\mu P_\mu$ with eigenvalue $s$ or $s_R$ by
\begin{equation}
\{\psi^-_{[s,j]}\} = \Phi_+^\times([s,j]), 
\,\, \mbox{and} \,\,
\{\psi^G_{[s_R,j]}\} = \Phi_+^\times([s_R,j]).
\end{equation}
	The spaces $\Phi_+^\times([s,j])$ and $\Phi_+^\times([s_R,j])$ of generalized eigenvectors of $P^\mu P_\mu$ with eigenvalue $s_R$ and $s$ respectively have been formed with the kets $\ket{[s,j]\vel{p}j_3\,^-}$ and $\ket{[s_R,j]\vel{p}j_3\,^-}$ by 
Eqs. \eqref{eq:rel-gamow-state} and \eqref{eq:psi-lsket} 
in the same way as the representation spaces of an unitary irreducible representation of the Poincar\'{e} group $[m^2,j]$ have been formed with the Wigner basis vectors $\ket{[m^2,j]\vel{p}j_3}$ for every fixed real $m$.
	But the $\ket{[m^2,j]\vel{p}j_3}$ are ordinary Dirac kets (functionals on the Schwartz space) and are denoted by $\Phi^\times ([m^2,j])$ and the $\ket{[s,j]\vel{p}j_3\,^-}$ are Lippmann-Schwinger kets.

	Remarkably the Lippmann-Schwinger kets of 
Eq.\ \eqref{eq:rel-basis-vect} 
and the Gamow kets 
\eqref{eq:rel-gamow-vect}, 
when mathematically defined as functionals on Hardy spaces $\Phi_+$, do not span a unitary representation space of the whole
        Poincar\'{e} group
\begin{align}
\label{eq:poincaregroup}
\mathcal{P} 
= 
&\{(\Lambda, x) | \Lambda \in \overline{SO(3, 1)},
\nonumber
\\
&\qquad \det \Lambda= +1, \Lambda_0^0 \geq 1, x \in \mathbb{R}_{1,3}\},
\end{align}
but only span a representation space of a Poincar\'{e} {\it semigroup}:
\begin{eqnarray}
\label{eq:poincaresemigroup}
\mathcal{P}_+ 
= 
\{(\Lambda, x) | \Lambda \in \overline{SO(3, 1)}, \det \Lambda= +1, \Lambda_0^0 \geq 1,
\nonumber
\\
x \in \mathbb{R}_{1,3},\,\,
x^2= t^2-\mathbf{x}^2 \geq 0, t\geq 0 \}.
\end{eqnarray}
This semigroup consists of all proper orthochronous Lorentz transformations and of space-time translations in the forward light cone.
	This restriction to the forward light cone is an expression of Einstein's causality 
\cite{ref:34}.


	The reason for this is that the extension $U^\times(\Lambda,x) \supset  U^\dagger(\Lambda,x)$ of the unitary operator $U^\dagger(\Lambda,x)$ in $\mathcal{H} \subset \Phi_+^\times$ to the operator $U^\times(\Lambda,x)$ in $\Phi_+^\times$ cannot be defined for $(\Lambda,x)$ outside of $\mathcal{P}_+$, because the restriction $U(\Lambda,x)|_{\Phi_+}$ of $U(\Lambda,x)$ in $\mathcal{H}$ is not a bounded operator in the space $\Phi_+$.
	The transformation formulas of the Lippmann-Schwinger and Gamow kets 
\cite{ref:34} 
are otherwise very similar to Wigner's unitary transformations 
\cite{ref:30a}.

	The semigroup property of the Poincar\'{e} transformation of the Lippmann-Schwinger kets was a little surprising since it is a standard assumption that the interacting scattering states furnish a representation of the whole Poincar\'{e} group 
\cite{ref:33a}.

	Unless these in- and out- Lippmann-Schwinger states are mathematically defined, one cannot prove any transformation property at all.
	One could try to define them, like Dirac kets, as Schwartz space functionals with unitary group transformation property.

	But if the $\ket{[s,j]\vel{p}j_3\,^\mp}$ fulfill the Lippmann-Schwinger boundary conditions like  
Eq.\ \eqref{eq:ls} 
with the $\mp i \epsilon$, the $\mp i \epsilon$ prevents this and the $\ket{[s,j]\vel{p}j_3\,^\mp}$ {\it cannot} be given a mathematical meaning that allows transformations under the whole Poincar\'{e} group.
	Defined as functionals on the Hardy spaces $\Phi_\pm$, the $\ket{[s,j]\vel{p}j_3\,^-} \in \Phi_+^\times$ allow transformations under $\mathcal{P}_+$, and the $\ket{[s,j]\vel{p}j_3\,^+}\in \Phi_-^\times$ allow transformations under $\mathcal{P}_-$ (semigroup of the backward light cone)
\cite{ref:34}.
	Though this was surprising for the Lippmann-Schwinger kets, for the Gamow kets 
\eqref{eq:rel-gamow-vect}, 
one expected this kind of semigroup property from the time asymmetry $t\geq 0$ in 
Eq.\ \eqref{eq:timeevolution} 
of the non-relativistic Gamow vectors 
\eqref{eq:nonrel-gv}.
	The representations $[s_R,j]$ --- arrived at here from the pole definition of a resonance and the Hardy space axiom 
\eqref{eq:newaxiom} --- 
were contained in a classification of Poincar\'{e} semigroup representations 
\cite{ref:35}, 
where they were also advocated as candidates for unstable relativistic particles.

	It will now be shown that the Gamow states $\psi_{[s_R,j]}^G$ which are associated to the resonance pole position at a complex value $s_R$ have a well defined value of lifetime.
	This is in contrast to some statements in the literature (Ref.\ \onlinecite{ref:17} referring to Ref.\ \onlinecite{ref:41a}) that the question as to what is the lifetime of a relativistic unstable particle is not  meaningful, and that {\it only} the complex pole position $s_R$ is a physically meaningful entity as stated in Ref.\ \onlinecite{ref:41b}.
	We shall show that the ``lifetime in the rest frame'' is a uniquely defined quantity for all Gamow states $\psi_{[s_R,j]}^G$ of the representation space $[s_R, j]$.
	This lifetime is obtained from the transformation property of the Gamow kets under Poincar\'{e} transformations.
	It is given by $\tau= (-2 \hbar\, Im\sqrt{s_R})^{-1}$.

	In order to show this we need the general formula for the transformation of the Gamow kets 
under the Poincar\'{e} transformations $(\Lambda,x) \in \mathcal{P}_+$ which have been derived in 
Ref.\ \onlinecite{ref:34}. 
	The homogeneous Lorentz transformations $(\Lambda, x=0)$ are unitarily represented like in Wigner's representations.
	For our purpose here we need only the spacetime evolution which fulfills the forward light cone condition:
\begin{align}
(\Lambda, x)=(I,x=(t,\mathbf{x})),\,\, 
t\geq 0,\,\, t\geq \mathbf{x} \cdot \mathbf{v} 
\equiv 
\frac{\mathbf{r}}{c}\cdot \frac{\mbox{\boldmath $v$}}{c},
\label{eq:forward-lightcone-cond}
\end{align}
where $\mathbf{r}$ is the space translation and {\boldmath $v$} the velocity in regular units of [$m$] and [$m/s$] respectively.
	The spacetime translated Gamow ket as obtained in Ref.\ \onlinecite{ref:34} is given by
\begin{eqnarray}
&&U^\times(I,x)\ket{[s_R,j]\vel{p}j_3\,^-} 
\nonumber
\\
&&= e^{-i x \cdot P^\times} \ket{[s_R,j]\vel{p}j_3\,^-} 
\nonumber
\\
&&= e^{-i \gamma \sqrt{s_R}(t-\mathbf{x} \cdot \mathbf{v})} \ket{[s_R,j]\vel{p}j_3\,^-} 
\nonumber
\\
&&= e^{-i M_R \gamma (t-\mathbf{x} \cdot \mathbf{v})} e^{- (\Gamma_R/2) \gamma (t-\mathbf{x} \cdot \mathbf{v})} \ket{[s_R,j]\vel{p}j_3\,^-}, \quad
\label{eq:rel-translation}
\end{eqnarray}
and here $\mathbf{v}$ is the three-velocity of the decaying Gamow state:
\begin{subequations}
\begin{align}
&\vel{p}
=\gamma \mathbf{v}
=\frac{\mathbf{p}}{\sqrt{s_R}}
=\gamma \frac{\mbox{\boldmath $v$}}{c},
\\
&\gamma (\mathbf{v})
=\frac{1}{\sqrt{1-\mathbf{v}^2}}
=\sqrt{1+\vel{p}^2}
=\hat{p}^0.
\end{align}
\label{eq:velocities}
\end{subequations}
	In 
Eq.\ \eqref{eq:rel-translation}, 
we have used the parameterization 
\eqref{eq:masswidth-b}, 
$\sqrt{s_R}=M_R-i\Gamma_R/2$, because only for this parameterization is the mass in the phase factor and the width in the real exponential.

	The Born probability rate to detect the observable $\ketbra{\psi^-}{\psi^-}$ in the spacetime translated Gamow state is proportional to 
\begin{align}
&|\bra{\psi^-}U^\times(I,x)\ket{[s_R,j]\vel{p}j_3\,^-}|^2 
\nonumber
\\
& =
|\braket{U(I,x)\psi^-}{[s_R,j]\vel{p}j_3\,^-}|^2 
\label{eq:70c}
\end{align}
with
\begin{eqnarray} 
\mbox{$t\geq 0$ and $x^2=t^2-\mathbf{x}^2=t^2-r^2/c^2\geq 0$.}
\label{eq:70d}
\end{eqnarray}
The r.h.s of 
Eq.\ \eqref{eq:70c} 
represents the probability rate to detect the untranslated Gamow state with an observable which has been translated from $\ketbra{\psi^-}{\psi^-}$ into the forward light cone 
\eqref{eq:70d}.
	The l.h.s is the probability looked at from the Schr\"{o}dinger picture  and the r.h.s is the same looked at from the Heisenberg picture.
	The light cone condition 
\eqref{eq:70d} 
makes two statements:
\begin{enumerate}
\item A state needs to be prepared first (at $t=0$) before one can speak of probabilities for observables, and
\item probabilities cannot propagate with a velocity $r/t$ for which $t<r/c$ or $r/t>c$ (Einstein causality).
\end{enumerate}
	Using 
Eq.\ \eqref{eq:rel-translation} 
for the space time evolution of an unstable state with pole parameter $s_R$ 
and velocity $\mbox{{\boldmath $v$}}=c\mathbf{v}$, we obtain for the probability 
\eqref{eq:70c}
\begin{eqnarray}
&&|\bra{\psi^-} U^\times (I,x) \ket{[s_R,j]\vel{p}j_3\,^-}|^2
\nonumber 
\\
&&=
e^{-\Gamma_R \gamma (t-\mathbf{x} \cdot \mathbf{v})}|\braket{\psi^-}{[s_R,j]\vel{p}j_3\,^-}|^2.
\label{eq:rel-decayprob}
\end{eqnarray}
	Thus, the decay rate of a Breit-Wigner resonance with pole position $s_R$  obeys an exponential decay law with time dilation.
	If one uses the time $t'=\gamma(\mbox{\boldmath $v$})(t-\frac{\mathbf{r}\cdot\mbox{{\boldmath $v$}}}{c^2})$ in the rest frame of the decaying particle of velocity $\mbox{\boldmath $v$}=c\mathbf{v}=\frac{c}{\gamma}\vel{p}$, then from 
\eqref{eq:rel-decayprob}
\begin{eqnarray}
\mbox{Decay rate $\sim e^{-\Gamma_R t'}$}, \,\,\, 
t'=\gamma(\mbox{{\boldmath $v$}})
(1-\frac{\mathbf{r}\cdot \mbox{{\boldmath $v$}}}{c^2}).
\label{eq:rel-decayrate}
\end{eqnarray}
	This means the lifetime of a decaying relativistic resonance with pole position $s_R$ is a well defined property and its inverse is given by 
\begin{eqnarray}
\hbar/\tau = -2 Im \sqrt{s_R} \equiv \Gamma_R,
\label{eq:70g}
\end{eqnarray}
where $s_R$ is the complex pole position of the resonance pole.
\footnote{
The time of life of an individual trapped ion can also be observed and the lifetime defined by the exponential law 
\eqref{eq:rel-decayprob} 
turns out to be the ensemble average over these times of the life of the individual excited ion states 
\cite{ref:39a}.
}
	The decay rate 
\eqref{eq:rel-decayrate} 
is a probability and the lifetime defined by it is the property of an ensemble and not of an individual quantum systems.
	The lineshape of a resonance --- and therewith its width --- is also the property of an ensemble.
	the lifetime-width relation 
\eqref{eq:70g} 
is therefore a relation between statistical quantities and for an ensemble of decaying states and an ensemble of resonances.
	It cannot make a statement about individual quantum systems.
	Usually the ensemble of decaying states used for lifetime measurements is also an ensemble over a wide range of velocities 
\cite{ref:41c}, 
and the decay rate is measured as a function of the distance $z$ traveled with the velocity $v=\frac{z}{t}$ which is according to 
Eq.\ \eqref{eq:rel-decayprob} 
proportional to 
\begin{eqnarray*}
&&|\bra{\psi^-} U^\times (I, t=\frac{z}{v},0,0,z) \ket{[s_R,j]\vel{p}j_3\,^-}|^2
\\
&&=
e^{-\Gamma_R \frac{z}{\gamma v}} |\braket{\psi^-}{[s_R,j]\vel{p}j_3\,^-}|^2.
\end{eqnarray*}
	The $v$ of the unstable particle (e.g., $K^0$) is usually determined 
\cite{ref:41c} 
from the real momenta of the (supposedly) stable decay products (e.g., $\pi^+ \pi^-$) by momentum conservation, neglecting the small imaginary part of the momentum $\mathbf{p}=(M_R-i\Gamma_R/2)\frac{\gamma}{c}\bm{v}$.
	This may be a conceptual problem but not a practical one for the unstable particles for which $\tau=\hbar/\Gamma_R$ can be measured.

	The result 
\eqref{eq:rel-decayrate} 
could have been more easily obtained if one applied the time evolution  
$U^\times(I,(t,0))=e^{-i P^\times_o t} \equiv e^{-i H^\times t}$ to a Gamow ket at rest $\psi_{[s_R,j]}^G(0)=\ket{[s_R,j]\vel{p}=0,j_3\,^-}$.
	For this special case one obtains from 
Eq.\ \eqref{eq:rel-translation}  
%
\begin{eqnarray}
\psi_{[s_R,j]}^{G}(t)^- 
&\equiv&
e^{-i H^\times t} \ket{[s_R,j]\vel{p}=0,j_3\,^-} 
\nonumber
\\
&=&
e^{-i \sqrt{s_R}\, t} \ket{[s_R,j]\vel{p}=0,j_3\,^-}
\label{eq:rel-timeevolution}
\end{eqnarray}
for $t\geq 0$, where $s_R$ is the pole of the Breit-Wigner resonance in 
Eq.\ \eqref{association} or \eqref{eq:relbwamp}.
	From 
Eq.\ \eqref{eq:rel-timeevolution} 
we see that only for the parameterization of 
Eq.\ \eqref{eq:masswidth-b}, $\sqrt{s_R}=(M_R-i\Gamma_R/2)$, do we obtain the exponential law in the form
\begin{eqnarray}
\psi_{[s_R,j]}^G(t) = e^{-i M_R t} e^{-(\Gamma_R/2) t} \psi_{[s_R,j]}^G(0).
\label{eq:rel-timeevolution-masswidth}
\end{eqnarray}
	This means the lifetime of an unstable relativistic particle is a well defined quantity and does not depend upon the manner in which the decaying particle is prepared. 
	The lifetime is a property of the quantum mechanical state described by the space $\{\psi_{[s_R,j]}^G \}$ and the width is a property of the amplitude $a_{[s_R,j]}^{BW}(s)$ which by 
Eq.\ \eqref{association} 
corresponds to this space.
	Both describe a quantum mechanical ensemble and the lifetime width relation 
\eqref{eq:rel-decayrate} 
is a statement about ensemble parameters.


	From the derivation in 
Eqs. \eqref{eq:pole-1} and \eqref{eq:pole-2}, 
we see that the Gamow vector 
\eqref{eq:rel-gamow-vect} 
can only be obtained if we use for the resonance amplitude the Cauchy kernel 
\eqref{eq:relbwamp}.
	Therefore, $\Gamma_Z$ of 
Eq.\ \eqref{eq:omamp} 
is excluded and none of the other width parameters, e.g., $\bar{\Gamma}_Z$ of 
Eq.\ \eqref{eq:masswidth-a} 
will fulfill the lifetime-width relation.
	A well defined lifetime $\tau$ of an unstable relativistic state is precisely the inverse of a well defined width $\Gamma_R$ for the relativistic resonance characterized by $(M_R, \Gamma_R)$.

	The transformation property under causal Poincar\'{e} transformation of the Gamow state vector 
\eqref{eq:rel-gamow-vect} 
chooses $(M_R, \Gamma_R)$ as the mass and width definition of the $Z$-boson and other relativistic resonances.
	With this definition in 
Eq.\ \eqref{eq:amp} 
with 
Eq.\ \eqref{eq:relbwamp} 
one obtains, from the fits of lineshape (and asymmetries) of the $Z$-boson, the mass value of $Z^0$ as $M_R=91.1626\pm 0.0031 GeV$.
	This differs significantly from the on-the-mass-shell value $M_Z$ in 
Ref.\ \onlinecite{ref:20} 
and the Table \ref{tb:zboson}.
	The value $M_R$ also differs from the usual pole value $\bar{M}_Z$ defined by 
Eq.\ \eqref{eq:masswidth-a} with 
Eq.\ \eqref{eq:relbwamp}.

	For the hadron resonances the differences between the two pole values $\bar{M}_\Delta$ and $M_{\Delta R}$ is minimal.
	But for the $\rho$-resonance the difference between $\bar{M}_\rho$ and $M_{\rho R}$ are noticeable.
	It leads for instance to the discrepancy between the quoted values in Refs.\ 
\onlinecite{ref:22r} and \onlinecite{ref:22a} 
for the $\rho$-mass as remarked at the end of Sec.\ \ref{sec:2}.

\subsection{\label{subsec:4-5}The expansion of the scattering amplitude in terms of relativistic Breit-Wigner amplitudes as the counterpart to the complex basis vector expansion}

	In the analysis of the hadron data for the determination of the hadron masses in Table \ref{tb:hadrons}, one did not only use the resonance amplitude plus background 
\eqref{eq:amp} but also included a second resonance, e.g., the $\rho-\omega$ interference for the determination of $\bar{M}_\rho$ and $\bar{\Gamma}_\rho$ in Ref.\ 
\onlinecite{ref:22r}.
	This means 
Ref.\ \onlinecite{ref:22r} 
took for the scattering amplitude of $e^+e^-\rightarrow \rho \rightarrow  \pi^+\pi^-$ --- among many other formulas --- the ansatz:
\begin{eqnarray}
a_j(s)
&=&
\frac{r_\rho}{s-s_{R_1}}+ \frac{r_\omega}{s-s_{R_2}}+B_j(s)
\label{eq:rho-omega-amp}
\end{eqnarray}
for $|r_\omega/r_\rho| \ll 1$, as suggested by the heuristic formula 
\eqref{eq:multires-scattamp-w}.
	There is no theoretical justification in S-matrix theory for a formula like 
Eq.\ \eqref{eq:rho-omega-amp}.
	One either develops the S-matrix around the pole $s_{R_1}$ and obtains a Laurent expansion which is valid in a circle around $s_{R_1}$ with a radius that is smaller than the distance to the nearest pole $s_{R_2}$.
	Or one obtains a Laurent expansion around $s_{R_2}$ valid in a radius that does not include $s_{R_1}$. 
	Still 
Eq.\ \eqref{eq:rho-omega-amp} 
is the phenomenologically favored formula for two resonances in the same partial wave (also in nuclear physics 
\cite{ref:B78}).
	With the new hypothesis 
\eqref{eq:newaxiom} 
there is no problem to derive a formula like 
Eq.\ \eqref{eq:rho-omega-amp} 
as an equality between generalized functions.

	For this purpose we return to Eq.\ \eqref{eq:cont-deform}.
	The integral \eqref{eq:background} on the r.h.s of Eq.\ \eqref{eq:cont-deform} has been transformed in Ref.\ \onlinecite{ref:B76} into an integral over the scattering energies $m_0^2\leq s< \infty$ (using the van Winter theorem for Hardy functions), and one obtains for all $\psi^- \in \Phi_+$:
\begin{eqnarray}
\braket{\psi^-}{\phi^{bg}}
&=&
\int_{m_0^2}^\infty ds \braket{\psi^-}{s^-} \braket{^+s}{\phi^+}b_j(s).
\label{eq:rel-bg}
\end{eqnarray}
	Here $b_j(s)$ for $s\geq m_0^2$ is determined from $S_{II}(s)$ for $s \leq m_0^2$.
	It is a slowly varying function if there are no other singularities of the S-matrix $S_j(s)$ \cite{ref:B76}.
	This function $b_j(s)$ is different from zero and corresponds to the background amplitude $B_j(s)$ in 
Eq.\ \eqref{eq:amp}. 
	Precisely we choose, because of the convention in 
Eq.\ \eqref{eq:s-matrix-amp}, 
$b_j(s)=1+2iB_j(s)$ for the elastic and $b_j(s)=2iB_j(s)$ for the inelastic channels.
	If one inserts 
Eq.\ \eqref{eq:rel-bg} 
for 
Eq.\ \eqref{eq:background} 
into 
Eq.\ \eqref{eq:cont-deform}, 
inserts 
Eq.\ \eqref{eq:rel-s-matrix-element-simple} 
for the l.h.s. of 
Eq.\ \eqref{eq:cont-deform}, 
and uses 
Eq.\ \eqref{eq:pole-2} 
for the two pole terms on the r.h.s. of 
Eq.\ \eqref{eq:cont-deform}, 
then one obtains 
\begin{eqnarray}
&&\int_{m_0^2}^{\infty}ds \braket{\psi^-}{s^-}\braket{^+s}{\phi^+} S_j(s)
\nonumber
\\
&&=
\int_{m_0^2}^{\infty} ds \braket{\psi^-}{s^-}\braket{^+s}{\phi^+} b_j(s) 
\nonumber
\\
&&\quad+
\sum_{i=1}^{N=2} \int_{-\infty_{II}}^{+\infty} ds \braket{\psi^-}{s^-}\braket{^+s}{\phi^+} \frac{R^{(i)}}{s-s_{R_i}}.
\label{eq:bg-pole-final}
\end{eqnarray}
	This is an equation valid for all $\psi^- \in \Phi_+$ (all out-observables) and all $\phi^+ \in \Phi_-$ (all in-states).
	This means that 
Eq.\ \eqref{eq:bg-pole-final} 
is an equation valid for all $\braket{\psi^-}{s^-}\braket{^+s}{\phi^+} \in \mathbf{H}^2_{-} \cap \mathcal{S}|_{\mathbb{R}_+}$, i.e., for all functions which are products of Hardy functions in the lower half complex $s$ plane (second Riemann sheet).
	This is analogous to 
Eq.\ \eqref{eq:bg-and-pole}, 
which is an equation valid for all $\psi^- \in \Phi_+$ (all out-observables).
	Omitting the arbitrary $\psi^- \in \Phi_+$ in 
Eq.\ \eqref{eq:bg-and-pole} 
resulted in 
Eq.\ \eqref{eq:bg-and-pole-simple} 
as equality between kets in the space $\Phi_+^\times$ (functionals on $\Phi_+$), we write it again,
\begin{eqnarray}
\phi^+
=
\phi^{bg} + \sum_i \ket{s_{R_i}^-}c_{R_i}.
\label{eq:bg-pole-final-ez}
\end{eqnarray}
	In the same way omitting the arbitrary Hardy function $\bk{\psi^-}{s^-}\bk{^+s}{\phi^+}$ and the integrals in 
Eq.\ \eqref{eq:bg-pole-final}, one obtains the following equation between distributions:
\begin{equation}
      \theta(s-m_0^2) S_j(s)= \theta (s-m_0^2) b_j(s)+
     \sum_i \frac{R^{(i)}}{s-s_{R_i}}.
\label{eq:bg-pole-distribution}
\end{equation}

	Mathematically the two equations, 
Eqs.\ \eqref{eq:bg-pole-final-ez} 
and 
\eqref{eq:bg-pole-distribution}, 
are functional equations: 
Eq.\ \eqref{eq:bg-pole-final-ez} 
is a functional equation over the set of all $\psi^- \in \Phi_+$ and 
Eq.\ \eqref{eq:bg-pole-distribution} 
is a functional equation over the set of all test functions $\bk{\psi^-}{s^-}\bk{{}^+s}{\phi^-} \in \mathbf{H}^2_- \cap \mathcal{S}|_{\mathbb{R}_+}$.
	Physically 
Eq.\ \eqref{eq:bg-pole-distribution} 
expresses the ($j$-th partial) S-matrix element in terms of a background amplitude (the $B(s)$ in 
Eq.\ \eqref{eq:amp}) 
and a superposition of (interfering) Breit-Wigner resonance amplitudes, and 
Eq.\ \eqref{eq:bg-pole-final-ez} 
expresses the prepared in-state as a superposition of a non-exponential background vector and a superposition of exponentially evolving Gamow vectors.
	Each term in 
Eq.\ \eqref{eq:bg-pole-distribution} 
has a corresponding term in 
Eq.\ \eqref{eq:bg-pole-final-ez}; 
in particular to the non-resonant slowly varying background amplitude $b_j(s)=1+2iB_j(s)$ in 
Eq.\ \eqref{eq:bg-pole-distribution} 
corresponds the non-exponential background vector $\phi^{bg}$ in 
Eq.\ \eqref{eq:bg-pole-final-ez} and to each Breit-Wigner amplitude in 
Eq.\ \eqref{eq:bg-pole-distribution} corresponds a Gamow ket in Eq.\ \eqref{eq:bg-pole-final-ez}. 
	If we use 
Eq.\ \eqref{eq:s-matrix-amp}, 
the functional equation 
\eqref{eq:bg-pole-distribution} 
can also be written as 
\begin{equation}
      \theta(s-m_0^2) a_j(s)= \theta (s-m_0^2) B_j(s)+
     \sum_i \frac{r^{(i)}}{s-s_{R_i}}
\label{eq:bg-pole-distribution2}
\end{equation}
which is just the mathematically precise version of 
Eq.\ \eqref{eq:amp} 
with 
Eq.\ \eqref{eq:relbwamp} 
except that here we considered a second resonance poles.
	The sum in 
Eqs. \eqref{eq:bg-pole-distribution} and \eqref{eq:bg-pole-distribution2} 
can actually extend over a finite (or even infinite) number of poles.
	If one ignores the background vector $\phi^{bg}\rightarrow 0$ one arrives at the Weisskopf-Wigner approximation as used, e.g., for the $K^0$ system 
\cite{ref:B19} 
and extensively used in nuclear physics 
\cite{ref:55new}.

	If the background integral for $\braket{\psi^-}{\phi^{bg}}$ is taken along the negative real axis second sheet as done in 
Eq.\ \eqref{eq:background} 
then the first term on the r.h.s of 
Eq.\ \eqref{eq:bg-pole-final-ez} 
and the first term on the r.h.s of 
Eq.\ \eqref{eq:bg-pole-distribution2}, 
the background term, is exactly defined.
	The sums over $i$ in 
Eqs.\ \eqref{eq:bg-pole-final-ez} and \eqref{eq:bg-pole-distribution2} 
extend over {\it all} resonance poles in the $j$-th partial wave.
	But for particular applications one does not have to deform the contour in Fig.\ 1b all the way to the negative real axis (and then ignore it).
	From 
Eqs.\ \eqref{eq:bg-pole-final-ez} and \eqref{eq:bg-pole-distribution2}, 
one obtains a practical approximation method if one is only interested in the effect of a few resonances near-by.
	One deforms the contour of integration only passed these few resonances and consider the far away resonances as part of the background (which one may ignore).
	In this way one obtains a more practical Weisskopf-Wigner approximation which contains only the near-by resonances in the scattering amplitude 
\eqref{eq:bg-pole-distribution2} 
and in the prepared state 
\eqref{eq:bg-pole-final-ez}.

\section{\label{sec:5}Summary and conclusion}

	Quasistable particles are observed in two different ways, in the lineshape as function of (the center-of-mass) scattering energy and in the decay rate (or probability) as function of (rest-frame) time.
	The lineshape is mostly Lorentzian (Breit-Wigner) and the decay rate is mostly exponential.
	For the lineshape it is standard to splits the scattering amplitude into an idealized Breit-Wigner resonance amplitude $a^{\mathcal{R}}$ and some background, like in Eq.\ \eqref{eq:amp}.
	In contrast, for the decay rate, the overwhelming opinion has been that the decay of unstable particles is non-exponential 
\cite{ref:104, ref:wilkinson, ref:martin}.

	This opinion has its origin in a mathematical consequence 
\cite{ref:104} 
of the Hilbert space axiom 
\cite{ref:xx} 
of traditional quantum mechanics, i.e., of 
Eq.\ \eqref{eq:stdqm} 
with $\Phi=\mathcal{H}=\mbox{Hilbert Space}$ (complete with respect to the norm-topology).
	If one does not insist on a linear space with norm-convergence one has many more possibilities.
	One could choose for $\Phi$ the Schwartz spaces.
	Then one can mathematically define the well accepted Dirac kets as functionals on $\Phi$, as it is done if one cares about mathematics \cite{ref:cpt}.
	But these Dirac kets are also insufficient for a theory of scattering and decay.
	The many useful heuristic notions that had been introduced to describe (resonance) scattering and decay phenomena -- like the two in- and out- Lippmann Schwinger kets with $\pm i\epsilon$ 
\cite{ref:101, ref:23}, 
the time asymmetric boundary conditions 
\cite{ref:103}, 
the Gamow states with complex energy and exponential time evolution \cite{ref:B6} --- do not fit into the traditional framework of quantum mechanics based on axiom \eqref{eq:stdqm}.
	Neither does the standard formalism of relativistic quantum field theory which contains the same $i\epsilon$ in the propagator.

		These $i\epsilon$ in $s$ (or an $i\epsilon'$ in $p^0$ or in the non-relativistic $E$) require that the energy wave functions must be better than Schwartz functions; they must also be continuable into the upper or lower complex $s$-plane.
		This does not necessarily mean that they need to be smooth Hardy functions as we assert by axiom \eqref{eq:newaxiom} or by \eqref{eq:hardy-whole} or \eqref{eq:rel-newaxiom}.
 	Axiom \eqref{eq:newaxiom} or \eqref{eq:rel-newaxiom} is the mathematical idealization which we made to assure that the triplets of spaces in Eq.\ \eqref{eq:newaxiom} are Rigged Hilbert spaces (so that the Dirac formalism applies and the Dirac basis vector expansions \eqref{eq:lsket-expansion-whole} or \eqref{eq:lsket-expansion-whole-rel} hold as the nuclear spectral theorem). 
	But analyticity is commonly presumed.

	Thus Dirac formalism and the $\pm i \epsilon$ of the propagator suggest the Hardy space axiom \eqref{eq:newaxiom}.
	With the Hardy space axiom the (non-relativistic  and) relativistic interaction incorporating ``in-'' and ``out-'' plane wave states \cite{ref:33a} are given a mathematical meaning, they are the functionals $\ket{[s,j]\vel{p} j_3 n^\pm} \in \Phi_\pm^\times$.
	Since these kets are now mathematically defined one can apply mathematics to show that they do {\it not} furnish a unitary (Wigner) representation of the Poincar\'{e} transformations, as often assumed \cite{ref:33a}.
	But --- independently of whether the imaginary part of energy is infinitesimal or finite --- they furnish only a semigroup representation in the forward ($-$) and backward ($+$) light cone respectively \cite{ref:34}.  
	This is the relativistic analogue of the time-evolution semigroup solutions for the Schr\"{o}dinger or Heisenberg equation which follow from Eq.\ \eqref{eq:newaxiom} in the non-relativistic case \cite{ref:28}.
	The time asymmetry given by the semigroup may be disturbing until one realizes that it expresses Einstein causality of the Born probabilities \cite{ref:34}.

		In relativistic quantum field theory the $i\epsilon$ rule of the propagator is a consequence of the assumption that the local (anti) commutator for space like separations vanishes (``local commutativity''), believed to be an expression of ``microscopic causality''.
		In our theory the $i\epsilon$ suggested the new axiom \eqref{eq:newaxiom} from which the semigroup and therewith Einstein causality follows as a mathematical result
\footnote{
From the connection between local commutativity and the semigroup representations of the Poincar\'{e} transformations via the $i\epsilon$ rule one can also understand why a system of axioms that contains the Poincar\'{e} {\it group} representation and local commutativity \cite{ref:cpt} would probably be condemned to triviality.
}.
	But the new axiom \eqref{eq:newaxiom} led to further conclusions which go far beyond the infinitesimal imaginary part $\pm i\epsilon$ in energy.

	The Lippmann-Schwinger kets \eqref{eq:rel-basis-vect} can be analytically continued into the entire complex semiplane (except at singularities) and the contour of integration for the S-matrix element (Born probability amplitude) \eqref{eq:rel-s-matrix-element} and \eqref{eq:rel-s-matrix-element-simple} can be deformed as shown in Fig.\ 1.
	The pole term \eqref{eq:pole-1} is then related to a functional \eqref{eq:rel-gv-func} and a Gamow ket \eqref{eq:rel-gamow-vect}, which associates to the relativistic Breit-Wigner amplitude \eqref{eq:relbwamp} an irreducible representation space of the Poincar\'{e} semigroup in the forward light cone, \eqref{association} and \eqref{eq:rel-gamow-state}.
	The Gamow kets \eqref{eq:rel-gamow-vect} are the {\it singularities} of the analytically continued scattering states, but they are not analytic continuations of scattering states or a continuous superposition thereof.
	In addition to the resonance states $\psi^G$ there is the background continuum, $\phi^{bg}$, \eqref{eq:background} and \eqref{eq:rel-bg} corresponding to the background term $B(s)$ in the scattering amplitude.
	Possible major or minor deviations from the exponential decay law 
are described by $\phi^{bg}$, whereas the Gamow state $\psi^G$ has purely exponential space-time evolution \eqref{eq:rel-decayprob}.

	The new mathematical theory thus establishes an exact correspondence 
between a Breit-Wigner amplitude for each resonance pole and the Gamow vector, $a^{BW} \Longleftrightarrow \psi^G$, and between the background amplitude $B$ and the vector $\phi^{bg}$.
	This is expressed for the relativistic case by the term by term correspondence between Eq.\ \eqref{eq:bg-pole-final-ez} and Eq.\ \eqref{eq:bg-pole-distribution2}.
	The $a^{BW} \Leftrightarrow \psi^G$ describes the quasistable particle per se, and the Breit-Wigner lineshape and the exponential decay are just two different manifestations of the same physical entity, the quasistable relativistic particle.

	In the non-relativistic case the exact form of the amplitude describing the resonance per se was never in doubt, it was given by the Lorentzian \eqref{eq:bwamp}, and from it the Gamow vector \eqref{eq:nonrel-gv} had been obtained \cite{ref:28}.
	The exact lifetime-width relation $\tau=\hbar/\Gamma$ of Eq.\ \eqref{eq:timeevolution} 
is a direct consequence of the new axiom \eqref{hardy-wave} and \eqref{eq:hrgs}.

	For the relativistic case one did not have a Weisskopf-Wigner approximation to go by, and the predominant opinion had been that relativistic resonances should not be characterized by two parameters like $(M,\Gamma)$ but had a complicated line shape and an energy dependent width $\Gamma(s)$.
	But since different hadrons of the {\it same} multiplet could have values for the width that varied by orders of magnitude (e.g., the $\Omega^-$ and the $\Delta$ in the decouplet), the idea that two real parameters $(M,\Gamma)$ or one complex parameter $s_R$ characterizes the relativistic quasistable states was never completely abandoned.
	When one noticed that the complex pole was the only gauge parameter independent definition of the $Z$-boson (and $W$) mass 
\cite{ref:20g, ref:20f} 
the pole of the S-matrix at a complex value $s_R$ became the favorite choice for the definition of a relativistic resonance.
	This was discussed in Sec.\ \ref{sec:2}, where the relativistic Breit-Wigner amplitude 
\eqref{eq:relbwamp} 
was identified as the part of the relativistic partial wave amplitude that describes the resonance per se.
	Only this Cauchy kernel \eqref{eq:relbwamp} for the resonance --- together with axiom \eqref{eq:newaxiom} --- allows the construction of the Gamow vector \eqref{eq:rel-gamow-vect-simple}.

	The resonance amplitude 
defines only the complex value $s_R$, not a mass $M$ and a width $\Gamma$ separately.
	From this it had been concluded in the past that the real and imaginary parts of $s_R$ separately have no physical significance \cite{ref:41b} and any of the parameterizations \eqref{eq:masswidth} should be equally valid.
	However using the exponential time evolution of the Gamow kets derived in Eq.\ \eqref{eq:rel-decayprob} we see that if the width is to be the total initial decay rate $\Gamma=\hbar R(0)$ then of the many possible parameterizations of $s_R$ one can only use Eq.\ \eqref{eq:masswidth-b} because only $\Gamma_R=\hbar/\tau$.

	In order to arrive at the conclusion \eqref{eq:rel-decayprob} we had to attribute to relativistic resonances the same spacetime properties as to relativistic stable particles, and define for each pole position $s_R$ of $S_j(s)$ relativistic Gamow kets 
\eqref{eq:rel-gamow-vect} 
which transform irreducibly under Poincar\'{e} transformations into the forward light cone.
	These relativistic Gamow vectors furnish a representation space of the causal Poincar\'{e} semigroup that is like the pole characterized by $[s_R, j]$.
	The vectors in this space $[s_R,j]$ of decaying states evolve exponentially 
\eqref{eq:70g}.
	Thus, each Gamow state $\psi_{[s_R,j]}^G$ given by Eq.\ \eqref{eq:rel-gamow-state} 
has a well defined lifetime \eqref{eq:70g} which is relativistically invariant and equal to the ``lifetime in the rest frame''.

	If there are two (or more) resonance poles in the same partial wave then the scattering amplitude contains a sum of the two (or more) Breit-Wigner amplitudes 
\eqref{eq:bg-pole-distribution}.
	This is what field theory for stable particles would suggest but it cannot be derived from the analyticity of the S-matrix alone. 
	The derivation requires the new hypothesis 
\eqref{eq:newaxiom} or equivalently the Hardy property of the functions in 
Eqs.\ \eqref{eq:hardy-whole}. 
	Corresponding to the sum of two (or more) Breit-Wigner amplitudes 
\eqref{eq:bg-pole-distribution} 
one derives a superposition of two (or more) interfering Gamow vector 
\eqref{eq:bg-pole-final-ez} for the prepared state. 
	Resonances are also observed in production processes \eqref{eq:production2}, 
	Gamow vectors therefore also emerge as intermediate states of production amplitudes \cite{ref:production1, ref:production2}.

	The background amplitude $B_j(s)$ in Eq.\ \eqref{eq:bg-pole-distribution2}, describes the ``contact terms'' \cite{ref:17} for direct production of the final state that does {\it not} go through resonance formation 
as, e.g., given by 
Eq.\ \eqref{nonresonant}.
	To this background amplitude $B_j(s)$ in Eq.\ \eqref{eq:bg-pole-distribution2}  corresponds a background vector $\phi^{bg}$ in the complex basis vector expansion 
\eqref{eq:bg-pole-final-ez}.
	This background vector is a continuous superposition 
of Lippmann-Schwinger scattering states \eqref{eq:rel-bg}.

	The approximation in which the background continuum 
\eqref{eq:rel-bg} 
is neglected, is a Weisskopf-Wigner approximation.
	In this approximation the scattering amplitude is a (finite) superposition of Breit-Wigner resonances, and the prepared state is a (finite) superposition of Gamow vectors, both of which have been well documented experimentally.

	To obtain all these new results one had to pay a price.
	This is the new hypothesis 
\eqref{eq:newaxiom} 
which requires that the energy distributions in the prepared state and the energy resolutions of the detected observables are described by much nicer energy wave functions \eqref{eq:hardy-whole}, than the Lebesgue-square integrable functions of Hilbert space quantum mechanics.
	As far as the preparation apparatus (accelerator) and the registration apparatus (detector) are concerned, the hypothesis that Eq.\ \eqref{eq:hardy-phi} is analytic in the lower complex semiplane and that Eq.\ \eqref{eq:hardy-psi} is analytic in the upper complex semiplane is just another acceptable mathematical idealization (because the apparatuses can probably not distinguish between a smooth function and a smooth function that can be analytically continued into the complex semiplane).
	But a mathematical theorem (Paley-Wiener theorem, see appendix of 
Ref.\ \onlinecite{ref:28}) 
leads to different asymmetric time dependence for the Fourier transforms of the two different kinds of Hardy function.
	Like the Stone-von Neumann theorem for the unitary group evolution of the Hilbert space the Paley-Wiener theorem is the mathematical underpinning for the time asymmetric semigroup evolution \eqref{eq:forward-lightcone-cond}, \eqref{eq:rel-translation}, and, in general, for the semigroup representations of the spacetime transformations.
	It follows as a mathematical consequence from the new Hardy space axiom \eqref{eq:newaxiom} \cite{ref:34}, which is the only modification of the standard axioms needed to obtain a consistent time asymmetric quantum theory that incorporates causality and many popular heuristic concepts.

	The analyticity in energy is the property of the Hardy function that is needed to unify the theoretical description of scattering resonances and decaying states and to explain such heuristic notions like Lippmann-Schwinger kets, Gamow vectors, Breit-Wigner amplitudes, and their interrelation.
	For the relativistic case it leads to a unique definition of resonance mass and of resonance width, which for the $Z$-boson gives the mass value \eqref{eq:mass-width-r}, which is none of the two quoted in Ref.\ \onlinecite{ref:20}.

\begin{acknowledgements}
	We had many discussions and received valuable advises and suggestions from many colleagues: D.\ Dicus, S.\ Weinberg, P.\ Kielanowski, G.\ Lopez Castro, J.\ Pestieau, G.\ H\"{o}hler, W.\ Hollik, M.\ Gadella, and S.\ Wickramasekara.
	We are particularly grateful to H.\ Kaldass who provided the calculations for resonance production.
	This work was supported in part by US NSF (042 1936) and the Welch Foundation.
\end{acknowledgements}



\end{document}